\documentclass[12pt]{article}

\usepackage{latexsym}
\usepackage{amssymb,amsfonts,amsmath}
\usepackage{graphicx} 
\usepackage{indentfirst}
\usepackage{bbm}
\usepackage{amssymb}
\usepackage{verbatim}
\usepackage{amsmath, amsthm,amssymb}
\usepackage{mathrsfs}
\usepackage{hyperref}
\usepackage{amsfonts}
\usepackage{dsfont}
\usepackage{cite}
\usepackage{xcolor}
\usepackage[multiple]{footmisc}

\parskip=\medskipamount

\newcommand {\cD}{{\cal D}}

\newcommand {\cN}{{\cal N}}

\newcommand {\cP}{{\cal P}}

\newcommand {\cR}{{\cal R}}

\newcommand {\mfo}{{\mathfrak o}}
\newcommand {\mfp}{{\mathfrak p}}

\newcommand {\mfs}{{\mathfrak s}}

\newcommand {\mfA}{{\mathfrak A}}

\newcommand {\mfF}{{\mathfrak F}}

\newcommand {\mfS}{{\mathfrak S}}

\newcommand{\dsR}{{\mathbb R}}

\newcommand{\dsA}{{\mathbb A}}
\newcommand{\dsT}{{\mathbb T}}

\def\a{\alpha}

\def\b{\beta}
\def\c{\chi}
\def\d{\delta}
\def\e{\epsilon}
\def\ce{\varepsilon}
\def\h{\eta}
\def\f{\phi}
\def\g{\gamma}
\def\G{\Gamma}

\def\k{\kappa}
\def\l{\lambda}
\def\m{\mu}

\def\s{\sigma}

\def\y{\psi}

\def\Q{\Theta}

\def\W{\Omega}

\def\Y{\Psi}
\def\tr{{\rm tr}}
\def\rd{{\rm d}}
\def\ri{{\rm i}}

\newcommand{\hf}{\frac12}

\newcommand{\be}{\begin{equation}}
\newcommand{\ee}{\end{equation}}
\newcommand{\bea}{\begin{eqnarray}}
\newcommand{\eea}{\end{eqnarray}}
\newcommand{\non}{\nonumber}
\newcommand{\bm}[1]{\mbox{\boldmath$#1$}}
\newcommand{\id}{\mathds{1}}
\newcommand{\ii}{\mathrm{i}}
\newcommand{\ph}{\phantom}
\def\double #1{#1{\hbox{\kern-2pt $#1$}}}

\newcommand{\hodge}{{\star}}
\newcommand{\dd}{\mathrm{d}}
\newcommand{\ad}{{\dot{\alpha}}}                           
\newcommand{\bd}{{\dot{\beta}}}

\newif\ifdtup

\newcommand{\bsubeq}{\begin{subequations}}
\newcommand{\esubeq}{\end{subequations}}

\numberwithin{equation}{section}

\newcommand{\sSL}{\mathsf{SL}}

\newcommand{\sSO}{\mathsf{SO}}
\newcommand{\sU}{\mathsf{U}}

\newcommand{\sOSp}{\mathsf{OSp}}

\begin{document}

\begin{titlepage}
\begin{flushright}
September, 2025 \\
\end{flushright}
\vspace{5mm}

\begin{center}
{\Large \bf 
Nonlinear realisation approach to extended supergravity theories in three dimensions} 
\end{center}


\begin{center}

{\bf Jake C. Stirling} \\
\vspace{5mm}

\footnotesize{
{\it Department of Physics M013, The University of Western Australia\\
35 Stirling Highway, Perth W.A. 6009, Australia}}  
~\\
\vspace{2mm}
~\\
Email: \texttt{jake.stirling@research.uwa.edu.au}\\
\vspace{2mm}

\end{center}

\begin{abstract}
\baselineskip=5pt

\end{abstract}
\vspace{-10mm}
We elaborate on the nonlinear realisation approach to spontaneously broken supergravity in three dimensions presented in \href{https://arxiv.org/abs/2304.09506}{arXiv:2304.09506}. Using this approach we provide a novel derivation of $\cN$-extended supergravity, with and without a cosmological term. It corresponds to a St\"uckelberg-type extension of the following theories: (i) the $(p,q)$ anti-de Sitter (AdS) supergravity theories with $p+q=\cN$, $p\geq q\geq 0$ proposed by Ach\'ucarro and Townsend; and (ii) the $\cN$-extended Poincar\'e supergravity of Marcus and Schwarz. We also apply the approach to obtain a St\"uckelberg reformulation of the supersymmetric Lorentz Chern-Simons action for arbitrary $\cN$. In our construction, the pure supergravity actions (Poincar\'e and AdS) share the invariance under two different local $\cN$-extended supersymmetries. One of them acts on the Goldstini, while the other supersymmetry leaves the Goldstini inert. The $\cN$-extended supersymmetric Lorentz Chern-Simons action proposed in our setting shares the former supersymmetry, but differs in the one that leaves the Goldstini inert. The supersymmetry that acts on the Goldstini can be used to gauge them away, and then the resulting actions coincide with that given in the literature. 
\vspace{10mm}
\begin{flushright}
	\it Dedicated to the memory of D. V. Volkov\\
	 on the 100th anniversary of his birth
\end{flushright}

\vfill

\vfill
\end{titlepage}

\newpage
\renewcommand{\thefootnote}{\arabic{footnote}}
\setcounter{footnote}{0}

\tableofcontents{}
\vspace{1cm}
\bigskip\hrule

\allowdisplaybreaks


\section{Introduction} \label{Section1}

The method of nonlinear realisations is an elegant and systematic approach that can be used to construct field theories with spontaneously broken symmetries. Its general formalism has been well established, and we refer the reader to several pioneering works and review papers on this subject \cite{CWZ,CCWZ,Isham,SalamS1,Volkov,Ogievetsky,SalamS2,ISS1,ISS2,Zumino,IO,BKY,McArthur}. Remarkably, this approach produced some of the earliest results in both rigid and local supersymmetry. These include the Goldstino action proposed in \cite{VA,AV} and, most notably, the Volkov-Soroka approach to spontaneously broken local supersymmetry \cite{VS,VS2}, which has been particularly influential for the present paper. These authors gauged the $\cN$-extended super-Poincar\'e group in four dimensions (4D) and proposed a super-Higgs mechanism by constructing the $\cN=1$ supergravity action with nonlinearly realised local supersymmetry
(see \cite{Volkov3,Volkov2} for a review and \cite{BMST} for a critical analysis of the Volkov-Soroka construction and modern developments).

In 2021, Kuzenko revisited the Volkov-Soroka construction and showed that, for special choices of the parameters of the theory, the $\cN=1$ Volkov-Soroka action is invariant under two distinct local supersymmetries \cite{K2021}. One of these, present for arbitrary values of the parameters, acts nontrivially on the Goldstino and can be used to gauge away the Goldstino. The other supersymmetry emerges only in a special case and leaves the Goldstino inert. When the former supersymmetry is used to eliminate the Goldstino, the resulting action coincides with that proposed by Deser and Zumino for consistent supergravity in the first-order formalism \cite{DZ}. In this sense, pure $\cN=1$ supergravity is a special case of the Volkov-Soroka theory. This analysis provided a novel nonlinear realisation approach to constructing unbroken simple Poincar\'e supergravity theories in 4D. In appendix \ref{Appendix A} we derive a supplementary result that builds on \cite{K2021} by including a cosmological term in the supergravity action. 

Less than two years after \cite{K2021}, the present author and Kuzenko \cite{KS} extended the construction of \cite{K2021} to the case of 3D $\cN=1$ supergravity \cite{HT,HT2}, with and without a cosmological term. These results were also generalised to topologically massive $\cN=1$ supergravity \cite{DeserKay} and its cosmological extension \cite{Deser}. Taken together, this provided a lower-dimensional application of the ideas of Volkov and Soroka \cite{VS,VS2}, culminating in a nonlinear realisation approach to (cosmological) topologically massive $\cN=1$ supergravity in three dimensions.

In this paper, we will extend the construction of \cite{KS} to the case of $(p,q)$ anti-de Sitter (AdS) supergravities \cite{AT} and $\cN$-extended Poincar\'e supergravity \cite{MS}. We also consider the supersymmetric Lorentz Chern-Simons action for arbitrary $\cN$ \cite{LR,NG}.

It is worth noting that $\cN$-extended supergravity in three dimensions has been an active area of research for several decades. The action for $\cN$-extended Poincar\'e supergravity first appeared in \cite{MS} and was later obtained in \cite{AT} by taking the ``Poincar\'e limit" of a family of $\cN$-extended AdS supergravity theories. This family of theories is known as the $(p,q)$ AdS supergravity theories and were also constructed in \cite{AT}. They were formulated as Chern-Simons theories and are naturally associated with the 3D AdS supergroups $\sOSp(p|2;\dsR)\times\sOSp(q|2;\dsR)$. Superfield approaches to $\cN$-extended supergravity and its matter couplings were developed in several works, see e.g. \cite{GGRS,ZP88,KLT-M11,KT-M11}. In particular, the off-shell actions for (cosmological) topologically massive supergravity were presented in \cite{KLRSTM} for $\cN=2$ and in \cite{KN} for $\cN=3$ and $\cN=4$ (for $\cN=4$ see also \cite{KNS}). For each $\cN$, the locally supersymmetric Lorentz Chern-Simons terms used to formulate these topologically massive supergravity theories can be interpreted as the action for conformal supergravity. This was first done in \cite{vN} for $\cN=1$ and then\cite{RvN} for $\cN=2$ and finally \cite{LR,NG}  for arbitrary $\cN$. The action for $\cN$-extended conformal supergravity given in \cite{LR,NG} is on-shell for $\cN\geq 3$. Off-shell actions for $\cN=3,4,5$ were constructed in \cite{BKNT-M2} using the techniques developed in \cite{BKNT-M1} and for $\cN=6$ in \cite{NT,KNT}.  

This paper is organised as follows. In section \ref{Section2} we review the 3D analogue of the Volkov-Soroka construction that was first presented in \cite{KS}. Using this framework, we demonstrate in section \ref{Section3} that the action for $(p,q)$ AdS supergravity \eqref{AdS(p,q)} is invariant under two different local $\cN$-extended supersymmetries. One of them is present for arbitrary relative coefficients between the terms in \eqref{AdS(p,q)} and acts on the Goldstini, while the other supersymmetry emerges only in a special case and leaves the Goldstini invariant. The former can be used to gauge away the Goldstini, and then the resulting action coincides with the standard action for $(p,q)$ AdS supergravity in the first-order formalism. In subsection \ref{Subsection 3.2} we show that the same formalism of nonlinearly realised local supersymmetry can be used to describe $\cN$-extended Poincar\'e supergravity. By restricting to the $\cN=2$ case, in subsection \ref{Subsection3.3} we discuss as examples, $(1,1)$ and $(2,0)$ AdS supergravity that follow from the analysis performed in subsection \ref{subsection3.1}. In section \ref{Section4} we apply our approach to obtain a generalisation of the $\cN$-extended supersymmetric Lorentz Chern-Simons action \eqref{Stopo} and demonstrate that it is also invariant under two different local supersymmetries, but where the second one is distinct from the one presented in section \ref{Section3}. We argue in subsection \ref{Subsection4.1} that our approach is limited in its ability to generalise the analysis of section \ref{Section3} to topologically massive supergravity. The main body of the paper is accompanied by two technical appendices. Appendix \ref{Appendix A} extends the result given in \cite{K2021} to include a cosmological term. In appendix \ref{Appendix B} we collect the key formulae of the 3D two-component spinor formalism.

\section{Review of the Volkov-Soroka approach in three dimensions}\label{Section2}

In this section we give a recap of the Volkov-Soroka construction in 3D, following \cite{KS}. Let $\cP(3|p,q)$ be the three-dimensional $(\cN=p+q)$-extended super-Poincar\'e group with $p\geq q\geq 0$. Any element $g\in\cP(3|p,q)$ is a $(4|\cN)\times(4|\cN)$ supermatrix of the form{\footnote{Our parametrisation of the elements of $\cP(3|p,q)$ follows \cite{KPTv}, but where the $R$-symmetry group is $\sSO(p)\times\sSO(q)$.}}
\begin{subequations}
\begin{align}
		g&= g(b,\h,M,\cR) = s(b,\h)h(M,\cR)\equiv sh~,\\
		s(b,\h)&:=\left(
		\begin{array}{c|c|c}
			\id_2 & 0 & 0 \\
			\hline
			-\hat{b}+\frac{\ii}{2}\ce^{-1}\h^2& \id_2 & -\sqrt{2}\h^T \\
			\hline
			\ii\sqrt{2}\h & 0 & \id_\cN
		\end{array}
		\right) =\left(
		\begin{array}{c|c|c}
			\d_\a\,^\b & 0 & 0 \\
			\hline
			-b^{\a\b}+\frac{\ii}{2}\ce^{\a\b}\h^2& \d^\a\,_\b & -\sqrt{2}\h^\a\,_J \\
			\hline
			\ii\sqrt{2}\h_I\,^\b & 0 & \d_{IJ}
		\end{array}
		\right)~,\\
		h(M,\cR)&:= \left(
		\begin{array}{c|c|c}
			M & 0 & 0 \\
			\hline
			0 & (M^{-1})^T & 0 \\
			\hline
			0 & 0 & \cR
		\end{array}
		\right) =\left(
		\begin{array}{c|c|c}
			M_\a\,^\b & 0 & 0 \\
			\hline
			0 & (M^{-1})_\b\,^\a & 0 \\
			\hline
			0 & 0 & \cR_{IJ}
		\end{array}
		\right)\label{helem}~,
\end{align}
\end{subequations}
where $M\in \sSL(2,\dsR)$, $\cR\in \sSO(p)\times\sSO(q)\subset \sSO(\cN)$, 
$\eta = (\eta_I{}^\b)$, 
$\h^2:=\h_I^\a\h_{\a I}$, and $\hat b$ is defined in \eqref{B3.b}.\footnote{In the $p=\cN$, $q=0$ case we recover the $\cN$-extended super-Poincar\'e group defined in \cite{KS}, where it is understood that $\cP(3|\cN,0)\equiv\cP(3|\cN)$.}
The $\sSL(2,\dsR)$ invariant spinor metric $\ce=(\ce_{\a\b}) = -(\ce_{\b \a}) $ and its inverse $\ce^{-1}=(\ce^{\a\b}) = - (\ce^{\b\a}) $ are
defined in appendix \ref{Appendix B}. 
The group element $s(b,\h)$ is labelled by three bosonic real parameters $b^a$ and $2\cN$ fermionic real parameters $\h_I\,^\a=\h^\a\,_I\equiv\h_I^\a$. 

Let us introduce Goldstone fields $Z^A(x)=\left(X^a(x),\Q_I^\a(x)\right)$ for spacetime translations $(X^a)$ and supersymmetry transformations $(\Q_I^\a)$. They parametrise the homogeneous space ($\cN$-extended Minkowski superspace)
\begin{equation}
	\mathbb{M}^{3|2\cN}=\frac{\cP(3|\cN)}{\sSL(2,\dsR)\times\sSO(p)\times\sSO(q)}
\end{equation}
according to the rule:
\begin{equation}
	\mfS(Z)=\left(
	\begin{array}{c|c|c}
		\id_2 & 0 & 0 \\
		\hline
		-\hat{X}+\frac{\ii}{2}\ce^{-1}\Q^2& \id_2 & -\sqrt{2}\Q^T \\
		\hline
		\ii\sqrt{2}\Q & 0 & \id_{\cN}
	\end{array}
	\right).
\end{equation}


A gauge super-Poincar\'e transformation acts as 
\begin{equation}
	g(x):Z(x)\rightarrow Z'(x)~,\qquad g\mfS(Z)=\mfS(Z')h~,
\end{equation}
with $g=sh$. This is equivalent to the following transformations of the Goldstone fields:
\begin{subequations}
	\begin{align}\label{stransGoldstone}
		s(b,\h):\qquad \hat{X}'&=\hat{X}+\hat{b}+\ii(\h^{\rm{T}}\Q-\Q^{\rm{T}}\h)~,\\
		\Q'&=\Q+\h~\label{stransGoldstino},
	\end{align}
\end{subequations}
and
\begin{subequations}
	\begin{align}
		h(M,\cR):\qquad \hat{X}'&=(M^{-1})^{\rm{T}}\hat{X}M^{-1}~,\\
		\Q'&=\cR\Q M^{-1}~.
	\end{align}
\end{subequations}

Introduce a connection $\mfA=\dd x^m\mfA_m$ taking its values in the super-Poincar\'e algebra $\mfp(3|p,q)$,	
\begin{equation}\label{superconn}
	\mfA:=\left(
	\begin{array}{c|c|c}
		\hf \W & 0 & 0 \\
		\hline
		-\hat{e}& - \hf\W^{\rm{T}} & -\sqrt{2}\y^{\rm{T}} \\
		\hline
		\ii\sqrt{2}\y & 0 & r
	\end{array}
	\right)=\left(
	\begin{array}{c|c|c}
		\hf \W_\a\,^\b & 0 & 0 \\
		\hline
		-e^{\a\b}& - \hf \W^\a\,_\b & -\sqrt{2}\y^\a\,_J \\
		\hline
		\ii\sqrt{2}\y_I\,^\b & 0 & r_{IJ}
	\end{array}
	\right)~,
\end{equation}
and possessing the gauge transformation law
\begin{equation}
	\mfA'=g\mfA g^{-1}+g\dd g^{-1}~.
\end{equation}
Here the one-form $\W_\a\,^\b$ is related to the Lorentz connection $\W^{ab}=\dd x^m\W_m\,^{ab}=-\W^{ba}$ as
\begin{equation}
	\W_\a\,^\b=\frac{1}{2}\ce_{abc}(\g^a)_\a\,^\b\W^{bc}.
\end{equation}
As in the first-order formalism to gravity, the Lorentz connection is an independent field and may be expressed in terms of the other fields by requiring it to be on-shell. The one-form $e^{\a\b}$ is the spinor counterpart of the dreibein $e^a=\dd x^m e_m\,^a$. The fermionic one-forms $\y_I\,^\b$ describe $\cN$ gravitini. Finally, the one-form $r_{IJ} = -r_{JI}$ is the $\sSO(p)\times \sSO(q)$ gauge field written as a $(p+q)\times(p+q)$ block diagonal matrix,

\begin{equation}\label{SO(N) conn}
	r_{IJ}:=\left(
	\begin{array}{c|c}
		r_{\overline{I}\overline{J}} & 0 \\
		\hline
		0 & r_{\underline{I}\underline{J}}
	\end{array}\right)
\end{equation}
where we have introduced the notation
\begin{subequations}
	\begin{align}\label{so(p)so(q)connections}
		r_{\overline{I}\overline{J}}&=-r_{\overline{J}\overline{I}}~,\qquad \overline{I},\overline{J}=1,\ldots,p~,\\
		r_{\underline{I}\underline{J}}&=-r_{\underline{J}\underline{I}}~,\qquad \underline{I},\underline{J}=1,\ldots,q~,\\
		r_{\overline{I}\underline{J}}&=r_{\underline{I}\overline{J}}=0~.
	\end{align}
\end{subequations}
We identify the fields $r_{\overline{I}\overline{J}}$ and $r_{\underline{I}\underline{J}}$ as the $\sSO(p)$ and $\sSO(q)$ gauge fields, respectively.

It should be pointed out that our parametrisation of the super-Poincar\'e algebra follows \cite{BKNT-M2}\footnote{With the distinction that the $R$-symmetry subalgebra is taken to be $\mfs\mfo(p)\oplus\mfs\mfo(q)$.} and differs from \cite{KPTv}. Under an infinitesimal Lorentz transformation 
\begin{subequations}
	\bea
	\d x^a = \l^a{}_b x^b = \varepsilon^{abc} \l_b x_c~, \qquad \l_{ab} = -\l_{ba}
	\eea
	a two-component spinor $\psi_\a$ transforms as 
	\bea
	\d \psi_\a = \hf \l_\a{}^\b \psi_\b~, \qquad \l_{\a\b} = \l_{\b\a}~,
	\eea 
\end{subequations}
where the Lorentz parameters $\l_{ab}$, $\l_a$ and $\l_{\a\b}$ are related to each other according to the rules \eqref{B.11}, \eqref{B.12} and \eqref{B.13}.

Associated with $\mfS$ and $\mfA$ is the different connection
\begin{equation}
	\dsA:=\mfS^{-1}\mfA\mfS+\mfS^{-1}\dd\mfS~,
\end{equation}
with gauge transformation law
\begin{equation}\label{dsAtrans}
	\dsA'=h\dsA h^{-1}+h\dd h^{-1}~,
\end{equation}
for an arbitrary gauge parameter $g=sh$. This connection is the main object in the Volkov-Soroka construction. 
Direct calculations give the explicit form of $\dsA$
\begin{equation}
	\dsA:=\left(
	\begin{array}{c|c|c}
		\hf \W & 0 & 0 \\
		\hline
		-\hat{E} & -\hf \W^{\rm{T}} & -\sqrt{2}\Y^{\rm{T}} \\
		\hline
		\ii\sqrt{2}\Y & 0 & r
	\end{array}
	\right),
\end{equation}
where we have defined 
\begin{subequations}\label{Compositefields}
	\begin{align}
		\hat{E}&:=\hat{e}+\cD\hat{X}+\ii\left(\cD\Q^{\rm{T}}\Q-\Q^{\rm{T}}\cD\Q\right)+2\ii\left(\y^{\rm{T}}\Q-\Q^{\rm{T}}\y\right)~,\\
		\Y&:=\y+\cD\Q~,\qquad
		\Y^{\rm{T}}=\y^{\rm{T}}+\cD\Q^{\rm{T}}~,
	\end{align}
\end{subequations}
and $\cD$ denotes the covariant derivative,
\begin{subequations}\label{Covderivs}
	\begin{align}
		\cD\hat{X}&=\dd\hat{X}- \hf \hat{X}\W- \hf \W^{\rm{T}}\hat{X}~,\\
		\cD\Q&=\dd\Q-\hf \Q\W+r\Q~,\qquad
		\cD\Q^{\rm{T}}=\dd\Q^{\rm{T}}- \hf\W^{\rm{T}}\Q^{\rm{T}}-\Q^{\rm{T}}r~.
	\end{align}
\end{subequations}
Equation (\ref{dsAtrans}) is equivalent to the following gauge transformation laws:
\begin{subequations}
	\begin{align}
		\W'&=M\W M^{-1}+M\dd M^{-1}~,\\
		r'&=\cR r\cR^{-1}+\cR\dd\cR^{-1}
	\end{align}
\end{subequations}
and
\begin{subequations}
	\begin{align}
		\hat{E}'&=(M^{-1})^{\rm{T}}\hat{E}M^{-1}~,\\
		\Y'&=\cR\Y M^{-1}~.
	\end{align}
\end{subequations}
It is worth pointing out that the supersymmetric one-forms $E^a$ and $\Y_I\,^\b$ transform as tensors with respect to the Lorentz and $\sSO(p)\times\sSO(q)$ gauge groups.

Under a supersymmetry transformation, $g=s(0,\h)$, one can use the Goldstone field transformations (\ref{stransGoldstone}) and (\ref{stransGoldstino}) to deduce the local supersymmetry transformation laws of the gravitini and the dreibein
\begin{subequations}\label{SUSYelem}
	\begin{align}\label{SUSYelem1}
		\y'&=\y-\cD\h~,\\
		\hat{e}'&=\hat{e}+2\ii\left(\h^{\rm{T}}\y-\y^{\rm{T}}\h\right)+\ii\left(\cD\h^{\rm{T}}\h-\h^{\rm{T}}\cD\h\right)\label{SUSYelem2}~.
	\end{align}
\end{subequations}
In the infinitesimal case, these supersymmetry transformation laws take the form
\begin{subequations}\label{2.188}
	\begin{equation}
		\d_\h\y=-\cD\h~,\qquad\d_\h e^a=2\ii\,{\rm tr}(\h\g^a\y^{\rm{T}})~.
	\end{equation}
	These should be accompanied by the supersymmetry transformations of the Goldstone fields 
	\bea
	\d_\h X^a = -\ri {\rm tr} (\Theta \g^a \eta^{\rm T})~, \qquad \d_\h \Theta = \eta~.
	\eea
\end{subequations}

A local Poincar\'e translation is given by $g=s(b,0)$. It acts on the Goldstone vector field $X^a$ and the dreibein $e^a$ as follows
\begin{equation}\label{transgauge}
	X'^a=X^a+b^a~,\qquad e'^a=e^a-\cD b^a~,
\end{equation}
while leaving the Goldstini and gravitini inert.

The curvature tensor is found through 
\begin{equation}
	\dsR=\dd\dsA-\dsA\wedge\dsA~,\qquad \dsR'=h\dsR h^{-1}~.
\end{equation}
Direct calculations give
\begin{equation}
	\dsR:=\left(
	\begin{array}{c|c|c}
		\hf R & 0 & 0 \\
		\hline
		-\hat{\mathbb T}& -\hf R^{\rm{T}} & -\sqrt{2}\cD\Y^{\rm{T}} \\
		\hline
		\ii\sqrt{2}\cD\Y & 0 & F
	\end{array}
	\right)~,
\end{equation}
where $R=(R_\a\,^\b)$ is the Lorentz curvature, $F=(F_{IJ})$ is the Yang-Mills field strength,
\begin{align}\label{gravfieldstrength}
	\cD\Y&=\dd\Y- \hf \Y\wedge\W-r\wedge\Y~,\qquad
	\cD\Y^{\rm{T}}=\dd\Y^{\rm{T}}+ \hf \W^{\rm{T}}\wedge\Y^{\rm{T}}-\Y^{\rm{T}}\wedge r
\end{align}
are the gravitino field strengths, and 
\begin{equation}
	\hat{\mathbb T}=\dd\hat{E}- \hf\hat{E}\wedge\W+ \hf \W^{\rm{T}}\wedge\hat{E}-2\ii\Y^{\rm{T}}\wedge\Y=\cD\hat{E}-2\ii\Y^{\rm{T}}\wedge\Y
\end{equation}
is the supersymmetric torsion tensor. In vector notation, the torsion tensor reads
\begin{equation}
	\dsT^a=\cD E^a-\ii\Y\wedge\g^a\Y^{\rm{T}}~.
\end{equation}
The Lorentz curvature tensor with spinor ($R_\a{}^\b$) and vector ($R^a{}_b)$ indices has the form 
\bea
R_\a{}^\b = \rd \Omega_\a{}^\b - \hf \W_\a{}^\g \wedge \W_\g{}^\b ~, \qquad 
R^a{}_b = \rd \Omega^a{}_b -  \W^a{}_c \wedge \W^c{}_b ~.
\eea
Finally the Yang-Mills field strength reads
\be\label{YMFS}
F=\dd r-r\wedge r~.
\ee

Using the above results, one can construct a locally supersymmetric action. 
With the notation $E= \det (E_m{}^a)$,
gauge-invariant functionals include the following:
\begin{itemize}
	\item The Einstein-Hilbert action
	\begin{equation}\label{EHaction}
		S_{\text{EH}}=\frac{1}{2}\int\ce_{abc}E^a\wedge R^{bc} =\frac{1}{2}\int \rd^3 x \, E\, R~;
	\end{equation}
	\item The Rarita-Schwinger action 
	\begin{equation}\label{RSaction}
		S_{\text{RS}}=\ii\int\Y_I^\a\wedge\cD\Y_{\a I} =\ii\int \rd^3 x \, E\, \ce^{mnp}\Y_{m\,I}^{\phantom{m\,}\a}\cD_p\Y_{n\a I}~;
	\end{equation}
	\item The cosmological term
	\begin{equation}
		S_{\text{cosm}}=-\frac{1}{6}\int\ce_{abc}E^a\wedge E^b\wedge E^c
		= \int \rd^3 x\, E	~;
	\end{equation}
	\item The mass term
	\begin{equation}
		S_{\text{mass}}=\int\Y_I\wedge E^a\g_a\wedge\Y_I =\int \rd^3 x \, E\, \ce^{mnp}\Y_{mI}\g_n\Y_{pI}~; \label{mass}
	\end{equation}
	\item The Yang-Mills action
	\be
		S_{\text{YM}}=-\frac{1}{2}\int{\rm tr}(F\wedge\hodge F)=-\frac{1}{4}\int\dd^3x\,E\,{\rm tr}(F_{mn}F^{mn})~.
	\ee	
\end{itemize}
All of the above actions are invariant under the $R$-symmetry group $\sSO(p)\times\sSO(q)$. Making use of the $\sSO(p)\times\sSO(q)$ connection $r$ and the corresponding field strength $F$, we can construct standard Chern-Simons actions. In the $\cN=1$ case, a  linear combination of the functionals (\ref{EHaction})-(\ref{mass}) gives an action for spontaneously broken supergravity.


\section{A new local $\cN$-extended supersymmetry}\label{Section3}


In this section we generalise the pure $\cN=1$ supergravity (Poincar\'e and AdS) results of \cite{KS} to the $\cN$-extended case.


\subsection{$(p,q)$ anti-de Sitter supergravity theories}\label{subsection3.1}

In order to describe $(p,q)$ AdS supergravity theories, it is necessary to work with the $(\cN=p+q)$-extended super-Poincar\'e algebra, $\mfp(3|p,q)$. This is precisely the parametrisation described in section \ref{Section2} defined by the equations \eqref{superconn} and \eqref{SO(N) conn}.



We are going to show that the following special linear combination
\begin{equation}\label{AdS(p,q)}
	S_{\text{AdS}(p,q)}=S_{\text{SG}}+S_{\text{super-cosm}}+S_{\text{VCS}}~,
\end{equation}
where 
	\begin{align}
	S_{\text{SG}}&=S_{\text{EH}}-2S_{\text{RS}}=\frac{1}{2}\int\ce_{abc}E^a\wedge R^{bc}-2\ii\int\Y_I\wedge\cD\Y_I\label{SUGRA}~,\\
	S_{\text{super-cosm}}&=m^2S_{\text{cosm}}-\ii m_{IJ}(S_{\text{mass}})_{IJ}=-\frac{1}{6}m^2\int\ce_{abc}E^a\wedge E^b\wedge E^c\non\\
	&\qquad\qquad\qquad\qquad\qquad\qquad\quad\,-\ii m_{IJ}\int\Y_I\wedge E^a \wedge\g_a\Y_J\label{Ssupercosm}~,\\
	S_{\text{VCS}}&=\frac{1}{2m^2}m_{IJ}\int\Big(\dd r_{IK}\wedge r_{KJ}-\frac{2}{3}r_{IK}\wedge r_{KL}\wedge r_{LJ}\Big)\label{VCS}
\end{align}
and
\begin{equation}\label{massmatrix}
	m_{IJ}=m\,\text{diag}(\underbrace{1,\ldots,1}_{\text{p times}},\underbrace{-1,\ldots,-1}_{\text{q times}})~,\quad m\in\dsR
\end{equation}
possesses a new local $\cN$-extended supersymmetry described by the parameters $\e=(\e_I^\a)=(\e_{\overline{I}}^\a,\e_{\underline{I}}^\a)$. Making use of the $\cN$-extended supersymmetry transformation \eqref{2.188} and the local Poincar\'e translation (\ref{transgauge}) allows us to impose the unitary gauge
\bea
X^a =0~, \qquad \Theta_I^\a =0~.
\label{unitary}
\eea
Then \eqref{AdS(p,q)} turns into the action for $(p,q)$ AdS supergravity originally constructed in \cite{AT}. 

Under the new local $\cN$-extended supersymmetry, the composite fields $E^a$ and $\Y_I^\a$ are postulated to transform as: 
\begin{subequations} \label{N-extendedSUSY}
	\begin{equation}\label{CompSUSY}
		 \d_\e E^a=2\ii\e_I\g^a\Y_I~,\qquad \d_\e\Y_I^\a=-\cD\e_I^\a-\frac{1}{2}m_{IJ}(\e\g_a)_J^\a E^a~.
	\end{equation}
	The Goldstone fields are required to be  inert under this transformation,
	\begin{equation}\label{GoldstoneSUSY}
		\d_\e X^a=0~, \qquad \d_\e\Q_I^\a=0~.
	\end{equation}
	The elementary fields $\y_I^\a$ and $e^a$ transform as follows:
		\begin{align}
			\d_\e\y_I^\a&=-\cD\e_I^\a+\hf (\Q\d_\e\W)_I^\a-(\d_\e r\Q)_I^\a-\frac{1}{2}m_{IJ}(\e\g_a)_J^\a E^a~,\label{elemSUSY}\\
			 \d_\e e^a&=-\d_\e\W^{ab}X_b+2\ii\e_I\g^a\Y_I+2\ii\cD\e_I\g^a\Q_I-\frac{\ii}{4}\ce^{abc}\d_\e\W_{bc}\Q^2+\ii\d_\e r_{IJ}\Q_J\g^a\Q_I\non\\
			&\quad+\ii m_{IJ}(\e_J\g_b\g^a\Q_I)E^b~.\label{elemSUSY2}
		\end{align}
\end{subequations}
The dependence on $\d_\e\W$ and $\d_\e r_{IJ}$ in \eqref{elemSUSY} and \eqref{elemSUSY2} is such that the composite fields $E^a$ and $\Y_I^\a$ remain unchanged when the connections are displaced: $\W\rightarrow\W+\d_\e\W$ and $r_{IJ}\rightarrow r_{IJ}+\d_\e r_{IJ}$. It should be pointed out that this $\cN$-extended supersymmetry is analogous to the $\cN=1$ supersymmetry presented in \cite{KS}. As in the $\cN=1$ case, the 
transformation laws of $\W$ and $r_{IJ}$ will be determined by demanding the action \eqref{AdS(p,q)} to be invariant under this new local $\cN$-extended supersymmetry \eqref{N-extendedSUSY}. 

We now compute the corresponding variations of each action \eqref{SUGRA}-\eqref{VCS}. The total variation for the action \eqref{SUGRA} under the transformations \eqref{N-extendedSUSY} is functionally identical to the result in the $\cN=1$ case modulo terms arising from the presence of the connection \eqref{SO(N) conn},\footnote{The appearance of the Yang-Mills field strength arises from the Bianchi identity, $\cD\cD\Y=-\hf\Y\wedge R-\Y\wedge F$.}
\begin{align}\label{SUGRAvar}
	\d_\e S_{\rm SG}&= \hf \int\dsT^a\wedge\ce_{abc}\d_\e\W^{bc}-4\ii\int F_{IJ}\wedge\e_I^\a\Y_{J\a}-2\ii\int\Y_I^\a\wedge\Y_{J\a}\wedge\d_\e r_{IJ}\\\non
	&\quad+2\ii m_{IJ}\int\Big(\cD E^a\wedge(\e\g_a)_J^\a\Y_{I\a}+E^a\wedge(\cD\e\g_a)_J^\a\wedge\Y_{I\a}\Big)~.
\end{align}
The total variation of the action \eqref{Ssupercosm} under \eqref{N-extendedSUSY} reads
\begin{align}\label{Super-cosmvar}
	\d_\e S_{\rm {super-cosm}}&= -2\ii m_{IJ}\int E^a\wedge(\cD\e\g_a)_J^\a\wedge\Y_{I\a}\non\\
	&\quad+2m_{IJ}\int\Big(\Y_{I\a}\wedge\e_K^\a\Y_{K\b}\wedge\Y_J^\b+\Y_{I\a}\wedge\e_{K\b}\Y_K^\a\wedge\Y_J^\b\Big)~.
\end{align}
We highlight the survival of the terms quartic in fermions in (\ref{Super-cosmvar}), which is in contrast to the $\cN=1$ case considered in \cite{KS}. Before computing the variation of the action \eqref{VCS}, it will prove useful to first split it into its $\sSO(p)$ and $\sSO(q)$ counterparts as follows,
\begin{align}
	S_{\text{VCS}}&=\frac{1}{2m^2}m_{IJ}\int\Big(\dd r_{IK}\wedge r_{KJ}-\frac{2}{3}r_{IK}\wedge r_{KL}\wedge r_{LJ}\Big)\non\\
	&=\frac{1}{2m}\int\Big(\dd r_{\overline{IJ}}\wedge r_{\overline{JI}}-\frac{2}{3}r_{\overline{IJ}}\wedge r_{\overline{JK}}\wedge r_{\overline{KI}}\Big)\non\\
	&\quad-\frac{1}{2m}\int\Big(\dd r_{\underline{IJ}}\wedge r_{\underline{JI}}-\frac{2}{3}r_{\underline{IJ}}\wedge r_{\underline{JK}}\wedge r_{\underline{KI}}\Big)~.
\end{align}
Its corresponding variation is then
\begin{equation}\label{VCSvar}
	\d_\e S_{\rm VCS}=-\frac{1}{m}\int F_{\overline{IJ}}\wedge\d_\e r_{\overline{IJ}}+\frac{1}{m}\int F_{\underline{IJ}}\wedge\d_\e r_{\underline{IJ}}~,
\end{equation}
where $F_{\overline{I}\overline{J}}$ and $F_{\underline{I}\underline{J}}$ are the $\sSO(p)$ and $\sSO(q)$ field strengths, respectively.
Combining all variations \eqref{SUGRAvar}, \eqref{Super-cosmvar} and \eqref{VCSvar} gives
\begin{align}
	\d_\e S_{\text{AdS}(p,q)}&= \hf \int\dsT^a\wedge\ce_{abc}\d_\e\W^{bc}-4\ii\int F_{IJ}\wedge\e_I^\a\Y_{J\a}-2\ii\int\Y_I^\a\wedge\Y_{J\a}\wedge\d_\e r_{IJ}\non\\
	&\quad+2\ii m_{IJ}\int\cD E^a\wedge(\e\g_a)_J^\a\Y_{I\a}-\frac{1}{m}\int F_{\overline{IJ}}\wedge\d_\e r_{\overline{IJ}}+\frac{1}{m}\int F_{\underline{IJ}}\wedge\d_\e r_{\underline{IJ}}\non\\
	&\quad+2m_{IJ}\int\Big(\Y_{I\a}\wedge\e_K^\a\Y_{K\b}\wedge\Y_J^\b+\Y_{I\a}\wedge\e_{K\b}\Y_K^\a\wedge\Y_J^\b\Big)\non\\
	&=\hf \int\dsT^a\wedge\ce_{abc}\d_\e\W^{bc}-4\ii\int F_{IJ}\wedge\e_I^\a\Y_{J\a}-2\ii\int\Y_I^\a\wedge\Y_{J\a}\wedge\d_\e r_{IJ}\non\\
	&\quad+2\ii m_{IJ}\int(\dsT^a+\ii\Y_K\wedge\g^a\Y_K)\wedge(\e\g_a)_J^\a\Y_{I\a}-\frac{1}{m}\int F_{\overline{IJ}}\wedge\d_\e r_{\overline{IJ}}+\frac{1}{m}\int F_{\underline{IJ}}\wedge\d_\e r_{\underline{IJ}}\non\\
	&\quad+2m_{IJ}\int\Big(\Y_{I\a}\wedge\e_K^\a\Y_{K\b}\wedge\Y_J^\b+\Y_{I\a}\wedge\e_{K\b}\Y_K^\a\wedge\Y_J^\b\Big)~.
\end{align}
Using the identity \eqref{gammacontraction} and combining terms quartic in fermions yields
\begin{align}
	\d_\e S_{\text{AdS}(p,q)}&= \hf \int\dsT^a\wedge(\ce_{abc}\d_\e\W^{bc}+4\ii m_{IJ}\e_J\g_a\Y_I)-4\ii\int F_{IJ}\wedge\e_I^\a\Y_{J\a}\non\\
	&\quad-2\ii\int\Y_I^\a\wedge\Y_{J\a}\wedge(\d_\e r_{IJ}-4\ii m_{JK}\e_{[K}\Y_{I]})-\frac{1}{m}\int F_{\overline{IJ}}\wedge\d_\e r_{\overline{IJ}}\non\\
	&\quad+\frac{1}{m}\int F_{\underline{IJ}}\wedge\d_\e r_{\underline{IJ}}~.
\end{align}
Observe that if we set
\begin{equation}
	\d_\e r_{IJ}=4\ii m_{JK}\e_{[K}\Y_{I]}\quad\Longleftrightarrow\quad \d_\e r_{\overline{IJ}}=4\ii m\e_{[\overline{J}}\Y_{\overline{I}]}~,\quad \d_\e r_{\underline{IJ}}=-4\ii m\e_{[\underline{J}}\Y_{\underline{I}]}~,
\end{equation}
then we are left with,
\begin{equation}\label{AdS(p,q)var}
		\d_\e S_{\text{AdS}(p,q)}= \hf \int\dsT^a\wedge(\ce_{abc}\d_\e\W^{bc}+4\ii m_{IJ}\e_J\g_a\Y_I)~.
\end{equation}
This remaining variation vanishes provided
\begin{equation}
	\d_\e\W^{bc}=2\ii m_{IJ}\ce^{abc}\e_J\g_a\Y_I~. \label{Lorentzconnvar}
\end{equation}

Alternatively, we can work with a composite Lorentz connection obtained by imposing the constraint 
\begin{equation}\label{1.5constraint}
	\dsT^a=\cD E^a-\ii\Y_I\wedge\g^a\Y_I=\dd E^a+E^b\wedge\W^a\,_b-\ii\Y_I\wedge\g^a\Y_I=0~.
\end{equation}
In the case of vanishing Goldstone fields, $X^a=0$ and $\Q_I^\a=0$, one can uniquely solve \eqref{1.5constraint} for the connection giving its well-known expression in terms of the dreibein and gravitini, $\W=\W(e,\y)$. Moreover, by demanding that the constraint \eqref{1.5constraint} remains invariant under the transformations \eqref{N-extendedSUSY}, we can determine the unique second-order variation for the dual Lorentz connection $\W_{ma}:=\hf\ce_{abc}\W_m\,^{bc}$,
\begin{equation}\label{2ndLorentzconnvar}
	\d_\e\W_{ma}=-2\ii\e_I\Big(\g_m\mfF_{Ia}-\hf E_{ma}\g_b\mfF_I^b\Big)+\ii m_{IJ}\e_J\left(\ce_{abc}E_m\,^b\Y_I^c-\g_a\Y_{Im}\right)~,
\end{equation}
where
\bea\label{dualgrav}
\hodge\cD\Y_I=\dd x^m\mfF_{Im}~, \qquad \mfF_{Im}:=\frac{1}{2}\ce_{mnp}\mfF_I^{np}
\eea
are the Hodge duals of the gravitino field strengths \eqref{gravfieldstrength}
\bea\label{gravstrength}
\cD\Y_I=\frac{1}{2}\dd x^m\wedge\dd x^n\mfF_{Inm}~, 
\qquad \mfF_{Inm}:=\cD_n\Y_{Im}-\cD_m\Y_{In}=-\mfF_{Imn}~. 
\eea    

\subsection{The Poincar\'e supergravity limit}\label{Subsection 3.2}
In our approach, there are two ways to construct $\cN$-extended Poincar\'e supergravity. One way is to start with the connection \eqref{superconn}, but with the $\sSO(p)\times\sSO(q)$ gauge field \eqref{SO(N) conn} switched off and proceed as outlined in section \ref{Section2}. Then, one constructs the following functional
\be\label{PoincareSUGRA}
S_{\text{SG}}=S_{\text{EH}}-2S_{\text{RS}}=\frac{1}{2}\int\ce_{abc}E^a\wedge R^{bc}-2\ii\int\Y_I\wedge\cD\Y_I~.
\ee	
In the unitary gauge \eqref{unitary}, the functional \eqref{PoincareSUGRA} coincides with the action for $\cN$-extended Poincar\'e supergravity originally constructed in \cite{MS}. In addition to the first supersymmetry transformation \eqref{2.188}, it can also be shown that the action \eqref{PoincareSUGRA} possesses a second local $\cN$-extended supersymmetry defined by \eqref{N-extendedSUSY}, but with $m=0$. Specifically, one obtains the following variation, which is functionally identical to the $\cN=1$ case considered in \cite{KS},
\begin{equation}\label{PoincareSUGRAvar}
\d_\e S_{\text{SG}}=\hf \int\dsT^a\wedge\ce_{abc}\d_\e\W^{bc}~.
\end{equation}
This variation vanishes if $\d_\e\W^{bc}=0$. Alternatively, we can deal with a composite connection obtained by imposing the constraint \eqref{1.5constraint}, which makes the variation \eqref{PoincareSUGRAvar} vanish. 

Similar to the $\cN=1$ case, the variation \eqref{PoincareSUGRAvar} implies that the Volkov-Soroka approach can naturally lead to the 1.5 formalism \cite{TvN,CW}. In this formalism, the variation $\d_\e\W$ becomes irrelevant when the equation of motion of the Lorentz connection \eqref{1.5constraint} is satisfied. 

The second way to construct $\cN$-extended Poincar\'e supergravity is by taking the ``Poincar\'e limit" of the action \eqref{AdS(p,q)} as described in \cite{AT}. This involves first performing the rescaling $r_{IJ}\rightarrow r'_{IJ}=\frac{1}{m}r_{IJ}$ and then setting $m=0$. The action that remains is \eqref{PoincareSUGRA}, where there is no longer dependence on the $\sSO(p)$ and $\sSO(q)$ gauge fields. 

We emphasise that both approaches result in the functional \eqref{PoincareSUGRA}, which possesses two local $\cN$-extended supersymmetries. The first is the supersymmetry transformation \eqref{2.188}, while the second is defined by \eqref{N-extendedSUSY}, with $m=0$. In particular, although both approaches share the same Lorentz connection transformation law, $\d_\e\W^{bc}=0$, ensuring that \eqref{PoincareSUGRAvar} vanishes under the second supersymmetry \eqref{N-extendedSUSY}, this is not the case when considering the transformation law of the $\sSO(p)\times\sSO(q)$ connection under that same supersymmetry. In the first approach, the $\sSO(p)\times\sSO(q)$ connection is inert; it is excluded in the connection \eqref{superconn}. In the second approach, the $\sSO(p)$ and $\sSO(q)$ connections become absent after rescaling $r_{IJ}\rightarrow r'_{IJ}=\frac{1}{m}r_{IJ}$ and taking the limit $m\rightarrow 0$. However, there still remains a non-zero transformation law for them:
\begin{equation}
	\d_\e r'_{\overline{IJ}}=4\ii\e_{[\overline{J}}\Y_{\overline{I}]}~,\quad \d_\e r'_{\underline{IJ}}=-4\ii \e_{[\underline{J}}\Y_{\underline{I}]}~.
\end{equation}

It should be pointed out that both approaches lead to the action for $\cN$-extended supergravity where there is no distinction between $p$ and $q$ for fixed $\cN$ \cite{MS}. This action has been referred to as the $\cN=p+q$ extended Marcus-Schwarz Poincar\'e supergravity \cite{HIPT}. When either $p>1$ or $q>1$, there exists $(p,q)$ Poincar\'e supergravity theories with inherent distinction between $p$ and $q$, and these were first constructed in \cite{HIPT}. These authors explored a central extension of the $\cN$-extended super-Poincar\'e algebra by introducing central charges and their associated vector gauge fields.

\subsection{The $\cN=2$ case}\label{Subsection3.3}
In the $\cN=2$ case, there are two AdS supergravity theories: $(1,1)$ and $(2,0)$. Here we state these as examples of the general result described in subsection \ref{subsection3.1}.

\subsubsection{$(1,1)$ AdS supergravity}
When $p=q=1$, the $\sSO(p)$ and $\sSO(q)$ gauge fields are absent and consequently there is no vector Chern-Simons term in the action. The functional that remains is
\begin{align}\label{AdS(1,1)}
		S_{\text{AdS}(1,1)}&=\frac{1}{2}\int\ce_{abc}E^a\wedge R^{bc}-2\ii\int\Y_I\wedge\cD\Y_I-\frac{1}{6}m^2\int\ce_{abc}E^a\wedge E^b\wedge E^c\non\\
		&\quad-\ii m_{IJ}\int\Y_I\wedge E^a \wedge\g_a\Y_J~,
\end{align}
with $m_{IJ}=m\,\text{diag}(1,-1)$. In the unitary gauge \eqref{unitary}, the functional \eqref{AdS(1,1)} turns into the action for $(1,1)$ AdS supergravity \cite{AT}. One can introduce by hand an auxiliary vector field $A=\dd x^m A_m$ into the action \eqref{AdS(1,1)} in the following way
\begin{align}\label{AdS(1,1)aux}
	S_{\text{AdS}(1,1)}&=\frac{1}{2}\int\ce_{abc}E^a\wedge R^{bc}-2\ii\int\Y_I\wedge\cD\Y_I-\frac{1}{6}m^2\int\ce_{abc}E^a\wedge E^b\wedge E^c\non\\
	&\quad-\ii m_{IJ}\int\Y_I\wedge E^a \wedge\g_a\Y_J-\int\hodge A\wedge A~.
\end{align}
In the unitary gauge \eqref{unitary}, this action coincides with the action for type I supergravity derived in \cite{KLRSTM}\footnote{In the terminology of \cite{KT-M11}, the $(1,1)$ and $(2,0)$ AdS supergravity theories are referred to as type I and type II (minimal) supergravity with a cosmological term, respectively.}, with the distinction that the auxiliary complex scalar field $M$ introduced therein is on-shell. It satisfies the equation of motion $\bar{M}=-4\mu$, where $\m=\ii m/2$ and $|\m|^2$ is proportional to the cosmological constant. 

Integrating out the auxiliary field $A$ results in the on-shell action \eqref{AdS(1,1)}. The motivation behind the introduction of this auxiliary vector field will become more clear when we discuss topologically massive type I supergravity in subsection \ref{Subsection4.1}.

Below we show that it is possible to augment the new local supersymmetry \eqref{N-extendedSUSY} (for the case of $p=q=1$) by including a local supersymmetry transformation law for the auxiliary vector field $A$. We postulate its transformation law to be\footnote{The antisymmetric tensors $\ce^{IJ}=\ce_{IJ}$ are normalised as $\ce^{12}=\ce_{12}=1$.}
\begin{align}\label{ASUSY}
	\d_\e A&=-\ii\ce_{IJ}(\e_I\g_b\g_a\mfF_J^b)E^a-\ii A_a\e_I\g^a\Y_I-\ii A\e_I\g^a\Y_{Ia}+\ii\ce_{abc}A^b\e_I\Y_I^c E^a\non\\
	&\quad-\ii m_{IJ}\ce_{IK}\Y_J\e_K~.
\end{align}
If we denote 
\begin{equation}\label{S_A}
	S_A:=-\int\hodge A\wedge A~,
\end{equation}
and compute its variation under \eqref{ASUSY} we get 
\begin{align}
	\d_A S_A&=-2\ii\int\dd^3x\,E\big\{\ce_{IJ}A^a(\e_I\g_b\g_a\mfF_J^b)+A^aA_b(\e_I\g^b\Y_{Ia})+A^aA_a(\e_I\g^b\Y_{Ib})\big\}\non\\
	&\quad-2\ii m_{IJ}\int\dd^3x\,E\,\ce_{IK}A^a(\Y_{Ja}\e_K)~.
\end{align}
Due to the presence of the Hodge dual of $A$ in \eqref{S_A}, we must also compute the variation with respect to the dreibein $E^a$ \eqref{CompSUSY},
\begin{equation}
	\d_E S_A=2\ii\int\dd^3x\,E\big\{A^aA_a(\e_I\g^b\Y_{Ib})+A^aA_b(\e_I\g^b\Y_{Ia})\big\}~.
\end{equation}
The complete variation of the action \eqref{S_A} is then
\begin{equation}\label{S_Avar}
	\d_\e S_A=-2\ii\int\dd^3x\,E\,\ce_{IJ}A^a(\e_I\g_b\g_a\mfF_J^b)-2\ii m_{IJ}\int\dd^3x\,E\,\ce_{IK}A^a(\Y_{Ja}\e_K)~.
\end{equation}

In order to cancel this variation, we need to deform the transformation law for the composite field $\Y_I^\a$. Specifically we alter the $\Y_I^\a$ transformation \eqref{CompSUSY} by including the following $A$-dependent term
\begin{equation}\label{Psideformed}
\d_\e^{(A)}\Y_I^\a=-\frac{1}{2}\ce_{IJ}(\e_J\g_a\g_b)^\a A^b E^a~.
\end{equation}
Varying our auxiliary action \eqref{AdS(1,1)aux} under this additional variation \eqref{Psideformed} gives us
\begin{equation}\label{auxvar}
	\d_\Y^{(A)}S_{\text{AdS(1,1)}}=2\ii\int\dd^3x\,E\,\ce_{IJ}A^a(\e_I\g_b\g_a\mfF_J^b)+2\ii m_{IJ}\int\dd^3x\,E\,\ce_{IK}A^a(\Y_{Ja}\e_K)~.
\end{equation}
Summing the variations \eqref{S_Avar} and \eqref{auxvar} results in their cancellation and we stay with the variation computed earlier \eqref{AdS(p,q)var} for the case $p=q=1$.

\subsubsection{$(2,0)$ AdS supergravity}
If we take $p=2$ and $q=0$, there remains a single $\sSO(2)\cong\sU(1)$ gauge field, $r_{\overline{IJ}}:=\ce_{\overline{IJ}}A$, where $A$ is the $\sU(1)$ gauge field. The functional \eqref{AdS(p,q)} becomes
\begin{align}\label{AdS(2,0)}
	S_{\text{AdS}(2,0)}&=\frac{1}{2}\int\ce_{abc}E^a\wedge R^{bc}-2\ii\int\Y_{\overline{I}}\wedge\cD\Y_{\overline{I}}-\frac{1}{6}m^2\int\ce_{abc}E^a\wedge E^b\wedge E^c\non\\
	&\quad-\ii m\int\Y_{\overline{I}}\wedge E^a \wedge\g_a\Y_{\overline{I}}-\frac{1}{m}\int F\wedge A~,
\end{align}
where $F=\dd A$ is the $\sU(1)$ field strength and $m_{\overline{IJ}}=m\d_{\overline{IJ}}$. In the unitary gauge \eqref{unitary}, this functional turns into the standard on-shell action for $(2,0)$ AdS supergravity \cite{AT}. 
 
\section{Topological terms}\label{Section4}

 A unique feature of three dimensions is the existence of Chern-Simons terms that can be used to define topologically massive couplings
\cite{Siegel,JT,Schonfeld,DJT1,DJT2}. By setting $p=\cN$ and $q=0$ in the parametrisation defined by \eqref{superconn} and \eqref{SO(N) conn}, we can study a generalisation of the $\cN$-extended supersymmetric Lorentz Chern-Simons action \cite{LR,NG}  which involves the Goldstone fields $X^a$ and $\Q_I^\a$. The action is given by
\be\label{Stopo}
S_{\rm CSG}=S_{\rm LCS}+S_{\rm FCS}+S_{\rm VCS}~,
\ee
where
\begin{align}\label{SLCS}
	S_{\text{LCS}}&=\frac{1}{2}\int\tr\left(\W\wedge\dd\W-\frac{1}{3}\W\wedge\W\wedge\W\right)\nonumber\\
	&=\frac{1}{4}\int\dd^3x\,E\,\ce^{mnp}\left(\W_m\,^{ab}R_{npab}+\frac{2}{3}\W_{m\ph{a}b}^{\ph{m}a}\W_{n\ph{b}c}^{\ph{n}b}\W_{p\ph{c}a}^{\ph{p}c}\right)
\end{align}
is the Lorentz Chern-Simons term,
\begin{align}\label{SFCS}
	S_{\text{FCS}}&=2\ii\int\Big(\cD\Y_I^\a\wedge\hodge\cD\Y_{I\a}+\hodge\cD\Y_I^\a\wedge E_{\a\b}\wedge\hodge\cD\Y_I^\b\Big)\nonumber\\
	&=-2\ii\int\dd^3x\,E\,(\mfF_I^m\g^a\g^b\mfF_I^n)E_{mb}E_{na}
\end{align} 
 is the fermionic Chern-Simons term, and
 \begin{align}\label{SVCS}
 S_{\text{VCS}}&=-\int\Big(\dd r_{IJ}\wedge r_{JI}-\frac{2}{3}r_{IJ}\wedge r_{JK}\wedge r_{KI}\Big)\non\\
 &=\hf\int\dd^3x\,E\,\ce^{mnp}\,\tr\left(F_{mn}r_p-\frac{2}{3}r_mr_nr_p\right)
 \end{align}
 is the vector Chern-Simons term. The fermionic Chern-Simons term involves the gravitino field strengths \eqref{gravstrength} and their Hodge duals \eqref{dualgrav}. The vector Chern-Simons term is constructed from the $\sSO(\cN)$ connection \eqref{YMFS}. In the unitary gauge (\ref{unitary}), the functional (\ref{Stopo}) coincides with the $\cN$-extended supersymmetric Lorentz Chern-Simons action which is also known as the action for (on-shell) $\cN$-extended conformal supergravity \cite{LR,NG}.\footnote{We emphasise that while the functional \eqref{Stopo} is identical to that given in \cite{LR,NG}, it does not possess the symmetries consisting of dilatations and special conformal (bosonic and fermionic) transformations. This is simply due to the fact that our approach starts from the super-Poincar\'e group and not the superconformal group.}
 
 We now demonstrate that the action \eqref{Stopo} is invariant under a newly defined local $\cN$-extended supersymmetry described by the parameters $\a=(\a_I^\b)$. This supersymmetry acts on the composite fields $E^a$ and $\Y_I^\b$ in the following way:
\begin{subequations}
 \begin{equation}
  \d_\a E^a=2\ii\a_I\g^a\Y_I~,\qquad \d_\a\Y_I^\b=-\cD\a_I^\b-\frac{1}{2}\k(\a\g_a)_I^\b E^a~,\qquad\k\in\dsR~.
\end{equation}
 The Goldstone fields do not transform under this supersymmetry,
 \begin{equation}
 	\d_\a X^a=0~,\qquad\d_\a\Q_I^\b=0~.
 \end{equation}
The induced transformations on the elementary fields $\y_I^\b$ and $e^a$ are given by:
\begin{align}
	\d_\a\y_I^\b&=-\cD\a_I^\b+\hf (\Q\d_\a\W)_I^\b-(\d_\a r\Q)_I^\b-\frac{1}{2}\k(\a\g_a)_I^\b E^a~,\label{elemconfSUSY}\\
	\d_\a e^a&=-\d_\a\W^{ab}X_b+2\ii\a_I\g^a\Y_I+2\ii\cD\a_I\g^a\Q_I-\frac{\ii}{4}\ce^{abc}\d_\a\W_{bc}\Q^2+\ii\d_\a r_{IJ}\Q_J\g^a\Q_I\non\\
	&\quad+\ii \k(\a_I\g_b\g^a\Q_I)E^b~.\label{elemconfSUSY2}
\end{align}
 Finally we have following transformations of the gauge connections: 
 	\begin{align}
 	\d_\a\W_{ma}&=-2\ii\a_I\Big(\g_m\mfF_{Ia}-\hf E_{ma}\g_b\mfF_I^b\Big)+\ii \k\a_I\left(\ce_{abc}E_m\,^b\Y_I^c-\g_a\Y_{Im}\right)~,\label{LorentzconfSUSY}\\	
 	\d_\a r_{IJ}&=2\ii\a_{[J}\g^b\g^a\mfF_{I]b}E_a-2\ii \k\Y_{[J}\a_{I]}\label{SO(N)confSUSY}~.
 	\end{align}
 \end{subequations}
  Notably, the above transformation law for the Lorentz connection \eqref{LorentzconfSUSY} is functionally identical to that given in \eqref{2ndLorentzconnvar} for the $(\cN,0)$ case of the $(p,q)$ AdS supergravity theories presented in subsection \ref{subsection3.1}.
  
  As before, it is assumed that the dependence on $\d_\a\W$ and $\d_\a r_{IJ}$ in \eqref{elemconfSUSY} and \eqref{elemconfSUSY2} is such that the composite fields $E^a$ and $\Y_I^\b$ remain unchanged when the connections are perturbed: $\W\rightarrow\W+\d_\a\W$ and $r_{IJ}\rightarrow r_{IJ}+\d_\a r_{IJ}$.
 
Let us compute variations of the action (\ref{Stopo}) in parts, beginning with the variation of the Lorentz Chern-Simons term (\ref{SLCS}),
\begin{align}\label{SLCSvar} 
	\d S_{\text{LCS}}&=\ii\int\dd^3x\,ER(\a_I\g_a\mfF_I^a)+4\ii\int\dd^3x\,E\,G^{ab}(\a_I\g_b\mfF_{Ia})\non\\
	&\quad+2\ii \k\int\dd^3x\,E G^{ab}(\a_I\g_a\Y_{Ib})+4\k\int\dd^3x\,E(\Y_{Ja}\g_b\mfF_J^a)(\a_I\Y_I^b)~,
\end{align}
where $G_{ab}=R_{ab}-\frac{1}{2}\h_{ab}R$ is the Einstein tensor, $R_{ab}=R^{c}{}_{a{c}b}$ is the Ricci tensor and $R=-2\h^{ab} G_{ab}=\h^{ab} R_{ab}$ is the Ricci scalar. Varying the fermionic Chern-Simons term (\ref{SFCS}) with respect to $\Y_I^\a$ gives
\begin{align}\label{SFCSPsiVar}
	\d_\Y S_{\text{FCS}}&=-2\ii\int\dd^3x\,E\Big\{G^{ab}(\a_I\g_a\mfF_{Ib})+G^{ab}(\a_I\g_b\mfF_{Ia})+\frac{1}{2}R(\a_I\g_a\mfF_I^a)\non\\
	&\quad-2\ii(\Y_{Ia}\g^b\mfF_I^a)(\a_J\mfF_{Jb})+2F_{IJ}^a(\a_I\g_b\g_a\mfF_J^b)\Big\}\non\\
	&\quad-2\k\int\dd^3x\,E\Big\{2(\a_I \mfF_I^a)(\Y_{Jb}\g^b\Y_{Ja})+\ce^{abc}(\a_I\g_a\mfF_{Id})(\Y_{Jc}\g^d\Y_{Jb})\nonumber\\
	&\quad-\ce^{abc}(\a_I\g_d\mfF_I^d)(\Y_{Jc}\g_a\Y_{Jb})+\ce^{abc}(\a_I\g_d\mfF_{Ia})(\Y_{Jc}\g^d\Y_{Jb})\nonumber\\
	&\quad+\ii G^{ab}(\a_I\g_a\Y_{Ib})\Big\}\non\\
	&\quad+4\ii \k\int\dd^3x\,EF_{JI}^a(\a_I\Y_{Ja})~.
\end{align}
Combining the variations (\ref{SLCSvar}) and (\ref{SFCSPsiVar}) results in the cancellation of both the Ricci scalar curvature terms and the Einstein tensor terms proportional to $\k$ leaving\footnote{In deriving this variation, we have made use of the fact that the Hodge dual of the antisymmetric part $R_{[ab]}$ of the Ricci tensor can be expressed in terms of the fermionic fields via the first Bianchi identity in the presence of torsion.}
\begin{align}\label{combinedvar}
	\d S_{\text{LCS}}+\d_\Y S_{\text{FCS}}&=4\int\dd^3x\,E\Big\{(\mfF_I^b\g^a\Y_{Ib})(\a_J\mfF_{Ja})+\ce^{abc}(\mfF_{Id}\g_b\Y_I^d)(\a_J\g_c\mfF_{Ja})\non\\
	&\quad-\ii F_{IJ}^a(\a_I\g_b\g_a\mfF_J^b)\Big\}\non\\
	&\quad-2\k\int\dd^3x\,E\Big\{(\Y_{Ja}\Y_{Ib})(\a_I\g^b\mfF_J^a)+(\Y_{Ja}\g_b\Y_I^b)(\a_I\mfF_J^a)\nonumber\\
	&\quad+2(\Y_{Jb}\g^b\Y_{Ja})(\a_I\mfF_I^a)+\ce^{abc}(\Y_{Jc}\g_d\Y_{Jb})(\a_I\g_a\mfF_I^d)\nonumber\\
	&\quad-\ce^{abc}(\Y_{Jc}\g_a\Y_{Jb})(\a_I\g_d\mfF_I^d)+\ce^{abc}(\Y_{Jc}\g_d\Y_{Jb})(\a_I\g^d\mfF_{Ia})\non\\
	&\quad+\ce^{abc}(\Y_{Jd}\g_c\Y_{Ia})(\a_I\g_b\mfF_J^d)\Big\}\non\\
	&\quad+4\ii \k\int\dd^3x\,EF_{JI}^a(\a_I\Y_{Ja})~,
\end{align}
where the additional terms have arised from a Fierz rearrangement of the fourth term in (\ref{SLCSvar}).
We now vary the action (\ref{SFCS}) with respect to the composite field $E^a$. This variation reads
\begin{align}\label{ESFCSvar}
	\d_ES_{\text{FCS}}&=4\int\dd^3x\,E\big\{-(\mfF_I^a\mfF_{Ia})(\a_J\g^b\Y_{Jb})-\ce^{abc}(\mfF_{Ib}\g_c\mfF_{Ia})(\a_J\g_d\Y_J^d)\non\\
	&\quad+2(\mfF_I^a\mfF_I^b)(\a_J\g_b\Y_{Ja})+2\ce^{abc}(\mfF_I^d\g_c\mfF_{Ia})(\a_J\g_b\Y_{Jd})\big\}~.
\end{align}
The variation of $S_{\rm FCS}$ (\ref{SFCS}) with respect to the Lorentz connection is given by
\begin{align}\label{ConnSFCSvar}
	\d_\W S_{\text{FCS}}&=4\int\dd^3x\,E\Big\{\ce^{abc}(\mfF_I^d\g_d\Y_{Ic})(\a_J\g_b\mfF_{Ja})-\ce^{abc}(\mfF_{Ia}\g_d\Y_{Ic})(\a_J\g_b\mfF_J^d)\nonumber\\
	&\quad-\ce^{abc}(\mfF_I^d\g_a\Y_{Ic})(\a_J\g_b\mfF_{Jd})+(\mfF_I^a\Y_{Ib})(\a_J\g_a\mfF_J^b)\Big\}\non\\
	&\quad-2\k\int\dd^3x\,E\Big\{(\Y_J^b\g_b\mfF_J^a)(\a_I\Y_{Ia})+(\Y_{Ja}\g_b\mfF_J^a)(\a_I\Y_I^b)+(\Y_{Ja}\mfF_J^a)(\a_I\g^b\Y_{Ib})\nonumber\\
	&\quad-(\Y_{Jb}\mfF_J^a)(\a_I\g^b\Y_{Ia})-\ce^{abc}(\Y_{Jc}\mfF_{Jb})(\a_I\Y_{Ia})-\ce^{abc}(\Y_{Jc}\g_d\mfF_{Ja})(\a_I\g^d\Y_{Ib})\nonumber\\
	&\quad+\ce^{abc}(\Y_{Jc}\g_d\mfF_J^d)(\a_I\g_a\Y_{Ib})-\ce^{abc}(\Y_{Jc}\g_a\mfF_J^d)(\a_I\g_d\Y_{Ib})\Big\}~.
\end{align}
The final variation of $S_{\rm FCS}$ (\ref{SFCS}) is the contribution arising from the $\sSO(\cN)$ connection,
\begin{align}\label{AConnSFCSvar}
	\d_r S_{\text{FCS}}&=-8\int\dd^3x\,E\Big\{\ce^{abc}(\mfF_{Ia}\Y_{Jb})(\a_{[J}\mfF_{I] c})-(\mfF_{Ia}\Y_{Jb})(\a_{[J}\g^a\mfF_{I]}^b)\non\\
	&\quad+(\mfF_{Ia}\Y_{Jb})(\a_{[J}\g^b\mfF_{I]}^a)+(\mfF_I^a\g^b\Y_{Jb})(\a_{[J}\mfF_{I] a})-(\mfF_I^b\g^a\Y_{Jb})(\a_{[J}\mfF_{I] a})\non\\
	&\quad+\ce^{abc}(\mfF_{Ib}\g^d\Y_{Jd})(\a_{[J}\g_c\mfF_{I] a})-\ce^{abc}(\mfF_I^d\g_b\Y_{Jd})(\a_{[J}\g_c\mfF_{I] a})\Big\}\non\\
	&\quad-8\k\int\dd^3x\,E\Big\{\ce^{abc}(\Y_{Jc}\mfF_{Ia})(\a_{[I}\Y_{J]b})+(\Y_{Jb}\g^b\mfF_I^a)(\a_{[I}\Y_{J]a})\non\\
	&\quad-(\Y_{Ja}\g^b\mfF_I^a)(\a_{[I}\Y_{J]b})\Big\}~.
\end{align}
The variation of the vector Chern-Simons term (\ref{SVCS}) reads
\begin{equation}\label{SVCSvar}
	\d S_{\rm VCS}=4\ii\int\dd^3x\,EF_{IJ}^a(\a_I\g_b\g_a\mfF_J^b)-4\ii \k\int\dd^3x\,EF_{JI}^a(\a_I\Y_{Ja})~,
\end{equation}
which will cancel the terms proportional to the Yang-Mills field strength in (\ref{combinedvar}).

In order to show that the total variation of the action (\ref{Stopo}) vanishes, we adopt the strategy used in the $\cN=1$ case presented in \cite{KS}. In particular, we systematically perform Fierz rearrangements on the individual terms contained within the variations of (\ref{combinedvar}), (\ref{ConnSFCSvar}) and (\ref{AConnSFCSvar}). Our Fierz rearrangement procedure is divided into two parts: the first addresses terms that are not proportional to $\k$, while the second handles terms that are proportional to $\k$. For the terms independent of $\k$, we rearrange them into expressions of the form $(\mfF\mfF)(\a\Y)$, possibly with gamma-matrices wedged between the fields. For the $\k$-dependent terms, we rearrange them into forms like $(\Y\Y)(\a\mfF)$, again allowing for gamma-matrices to appear between fields. It is worth noting that the variation (\ref{ESFCSvar}) is already in this desired form and thus does not require any further Fierz manipulation. By applying the Fierz rearrangement rule for two-component spinors (\ref{Fierz}), we obtain the desired forms for the variations (\ref{combinedvar}), (\ref{ConnSFCSvar}) and (\ref{AConnSFCSvar}):
\begin{subequations}
	\begin{align}\label{Fierzvar1}
		\d S_{\text{LCS}}+\d_\Y S_{\text{FCS}}&=2\int\dd^3x\,E\Big\{-3(\mfF_I^b\mfF_{Ja})(\a_J\g^a\Y_{Ib})+(\mfF_{Ib}\g^a\mfF_{Ja})(\a_J\Y_I^b)\nonumber\\
		&\quad-\ce^{abc}(\mfF_I^d\g_c\mfF_{Ja})(\a_J\g_b\Y_{Id})-2\ii F_{IJ}^a(\a_I\g_b\g_a\mfF_J^b)\Big\}\non\\
		&\quad-2\k\int\dd^3x\,E\Big\{(\Y_{Ja}\Y_{Ib})(\a_I\g^b\mfF_J^a)+(\Y_{Ja}\g_b\Y_I^b)(\a_I\mfF_J^a)\nonumber\\
		&\quad+2(\Y_{Jb}\g^b\Y_{Ja})(\a_I\mfF_I^a)+\ce^{abc}(\Y_{Jc}\g_d\Y_{Jb})(\a_I\g_a\mfF_I^d)\nonumber\\
		&\quad-\ce^{abc}(\Y_{Jc}\g_a\Y_{Jb})(\a_I\g_d\mfF_I^d)+\ce^{abc}(\Y_{Jc}\g_d\Y_{Jb})(\a_I\g^d\mfF_{Ia})\non\\
		&\quad+\ce^{abc}(\Y_{Jd}\g_c\Y_{Ia})(\a_I\g_b\mfF_J^d)\Big\}\non\\
		&\quad+4\ii \k\int\dd^3x\,EF_{JI}^a(\a_I\Y_{Ja})~,
	\end{align}
	\begin{align}\label{Fierzvar2}
		\d_\W S_{\text{FCS}}&=2\int\dd^3x\,E\Big\{-(\mfF_I^b\g^a\mfF_{Ja})(\a_J\Y_{Ib})+(\mfF_I^b\mfF_{Ja})(\a_J\g^a\Y_{Ib})-2(\mfF_{Ia}\mfF_J^b)(\a_J\g^a\Y_{Ib})\non\\
		&\quad-2(\mfF_I^a\g^b\mfF_{Ja})(\a_J\Y_{Ib})+2(\mfF_I^a\mfF_{Ja})(\a_J\g^b\Y_{Ib})+2\ce^{abc}(\mfF_{Ia}\mfF_{Jb})(\a_J\Y_{Ic})\nonumber\\
		&\quad-\ce^{abc}(\mfF_I^d\g_b\mfF_{Ja})(\a_J\g_d\Y_{Ic})+2\ce^{abc}(\mfF_{Ib}\g_d\mfF_{Ja})(\a_J\g^d\Y_{Ic})\nonumber\\
		&\quad-\ce^{abc}(\mfF_I^d\g_d\mfF_{Ja})(\a_J\g_b\Y_{Ic})+\ce^{abc}(\mfF_{Ia}\g_b\mfF_J^d)(\a_J\g_d\Y_{Ic})+\ce^{abc}(\mfF_{Ia}\g_d\mfF_J^d)(\a_J\g_b\Y_{Ic})\non\\
		&\quad-\ce^{abc}(\mfF_{Ib}\g_c\mfF_{Jd})(\a_J\g_a\Y_I^d)\Big\}\non\\
		&\quad+2\k\int\dd^3x\,E\Big\{(\Y_{Ja}\g^a\Y_{Ib})(\a_I\mfF_J^b)-(\Y_{Ja}\Y_{Ib})(\a_I\g^a\mfF_J^b)\non\\
		&\quad+2(\Y_{Ja}\Y_I^b)(\a_I\g_b\mfF_J^a)+\ce^{abc}(\Y_{Ja}\g_c\Y_{Id})(\a_I\g_b\mfF_J^d)-\ce^{abc}(\Y_{Jc}\Y_{Ib})(\a_I\mfF_{Ja})\non\\
		&\quad-\ce^{abc}(\Y_{Jc}\g_d\Y_{Ia})(\a_I\g^d\mfF_{Jb})\Big\}~,
	\end{align}
\begin{align}\label{Fierzvar3}
	\d_r S_{\rm FCS}&=-4\int\dd^3x\,E\Big\{\ce^{abc}(\mfF_{Ia}\mfF_{Jc})(\a_I\Y_{Jb})+2(\mfF_{Ia}\mfF_I^b)(\a_J\g^a\Y_{Jb})-(\mfF_{Ia}\mfF_I^a)(\a_J\g^b\Y_{Jb})\non\\
	&\quad+\ce^{abc}(\mfF_I^b\g_c\mfF_I^d)(\a_J\g^a\Y_{Jd})-(\mfF_{Ia}\mfF_J^b)(\a_I\g^a\Y_{Jb})+(\mfF_{Ia}\mfF_J^a)(\a_I\g^b\Y_{Jb})\non\\
	&\quad-\frac{1}{2}\ce^{abc}(\mfF_{Ib}\g_c\mfF_J^d)(\a_I\g_a\Y_{Jd})+(\mfF_{Ia}\g^b\mfF_J^a)(\a_I\Y_{Jb})-(\mfF_I^b\mfF_{Ja})(\a_I\g^a\Y_{Jb})\non\\
	&\quad-\frac{1}{2}\ce^{abc}(\mfF_{Ib}\g_c\mfF_{Ia})(\a_J\g_d\Y_J^d)-\frac{1}{2}\ce^{abc}(\mfF_{Ib}\g_d\mfF_{Ia})(\a_J\g_c\Y_J^d)\non\\
	&\quad-\frac{1}{2}\ce^{abc}(\mfF_I^d\g_c\mfF_{Ja})(\a_I\g_b\Y_{Jd})+\frac{1}{2}\ce^{abc}(\mfF_{Ib}\g_c\mfF_{Ja})(\a_I\g_d\Y_J^d)\non\\
	&\quad+\frac{1}{2}\ce^{abc}(\mfF_{Ib}\g_d\mfF_{Ja})(\a_I\g_c\Y_J^d)\Big\}\non\\
	&\quad+2\k\int\dd^3x\,E\Big\{\ce^{abc}(\Y_{Jc}\g^d\Y_{Jb})(\a_I\g_d\mfF_{Ia})-\ce^{abc}(\Y_{Ic}\Y_{Jb})(\a_I\mfF_{Ja})\non\\
	&\quad-\ce^{abc}(\Y_{Ic}\g^d\Y_{Jb})(\a_I\g_d\mfF_{Ja})-2(\Y_{Ja}\g^b\Y_{Jb})(\a_I\mfF_I^a)\non\\
	&\quad-2\ce^{abc}(\Y_{Jd}\g_c\Y_{Ja})(\a_I\g_b\mfF_I^d)-(\Y_{Ia}\g^a\Y_{Jb})(\a_I\mfF_J^b)-(\Y_{Ia}\Y_{Jb})(\a_I\g^a\mfF_J^b)\non\\
	&\quad-\ce^{abc}(\Y_{Ia}\g_c\Y_{Jd})(\a_I\g_b\mfF_J^d)+(\Y_{Ia}\g^b\Y_{Jb})(\a_I\mfF_J^a)+(\Y_{Ib}\Y_{Ja})(\a_I\g^a\mfF_J^b)\non\\
	&\quad+\ce^{abc}(\Y_{Id}\g_c\Y_{Ja})(\a_I\g_b\mfF_J^d)\Big\}~.
\end{align}
\end{subequations}
Summing all variations (\ref{Fierzvar1}), (\ref{Fierzvar2}), (\ref{Fierzvar3}), (\ref{ESFCSvar}) and (\ref{SVCSvar}) gives
\begin{align}
	\d S_{\text{CSG}}&=\d S_{\text{LCS}}+\d_\Y S_{\text{FCS}}+\d_ES_{\text{FCS}}+	\d_\W S_{\text{FCS}}+\d_r S_{\rm FCS}+\d S_{\rm VCS}\nonumber\\
	&=2\int\dd^3x\,E\Big\{-\ce^{abc}(\mfF_I^d\g_b\mfF_{Ja})(\a_J\g_d\Y_{Ic})+2\ce^{abc}(\mfF_{Ib}\g_d\mfF_{Ja})(\a_J\g^d\Y_{Ic})\non\\
	&\quad-\ce^{abc}(\mfF_I^d\g_d\mfF_{Ja})(\a_J\g_b\Y_{Ic})+\ce^{abc}(\mfF_{Ia}\g_b\mfF_J^d)(\a_J\g_d\Y_{Ic})\non\\
	&\quad+\ce^{abc}(\mfF_{Ia}\g_d\mfF_J^d)(\a_J\g_b\Y_{Ic})-\ce^{abc}(\mfF_{Ib}\g_c\mfF_{Ia})(\a_J\g_d\Y_J^d)\non\\
	&\quad+2\ce^{abc}(\mfF_I^d\g_c\mfF_{Ia})(\a_J\g_b\Y_{Jd})+\ce^{abc}(\mfF_{Ib}\g_d\mfF_{Ia})(\a_J\g_c\Y_J^d)\non\\
	&\quad-\ce^{abc}(\mfF_{Ib}\g_c\mfF_{Ja})(\a_I\g_d\Y_J^d)-\ce^{abc}(\mfF_{Ib}\g_d\mfF_{Ja})(\a_I\g_c\Y_J^d)\Big\}\non\\
	&\quad+2\k\int\dd^3x\,E\Big\{\ce^{abc}(\Y_{Jc}\g_a\Y_{Jb})(\a_I\g_d\mfF_I^d)-\ce^{abc}(\Y_{Jc}\g_d\Y_{Jb})(\a_I\g_a\mfF_I^d)\non\\
	&\quad-2\ce^{abc}(\Y_{Jd}\g_c\Y_{Ja})(\a_I\g_b\mfF_I^d)\Big\}~.
\end{align}
In analogy to the $\cN=1$ case, this combination may be shown to be identically zero.  

\subsection{Topologically massive supergravity}\label{Subsection4.1}
In \cite{KS}, a nonlinear realisation approach was employed to demonstrate that the action for $\cN=1$ cosmological topologically massive supergravity is invariant under two different local supersymmetries. One of them acts on the Goldstino, while the other supersymmetry leaves the Goldstino invariant. The former can be used to gauge away the Goldstino, and then the resulting action coincides with that given in the literature \cite{Deser}. Here we argue why applying this approach to the $\cN=2$ case is inconsistent. For the $\cN=2$ case, there are two (minimal) topologically massive supergravity theories: topologically massive type I supergravity and topologically massive type II supergravity. 

\subsubsection{Topologically massive type I supergravity}
Topologically massive type I supergravity emerges from coupling the $\cN=2$ supersymmetric Lorentz Chern-Simons action \cite{RvN} with the off-shell $(1,1)$ AdS supergravity action \cite{KLRSTM}. In our framework, we have constructed the on-shell (auxiliary) component action for $(1,1)$ AdS supergravity \eqref{AdS(1,1)aux}. Although this action is not the off-shell one given in \cite{KLRSTM}, one can still couple it to the $\cN=2$ supersymmetric Lorentz Chern-Simons action since the scalar field $M$ remains auxiliary upon this coupling and can be integrated out at the end. Therefore, topologically massive type I supergravity \cite{KLRSTM} can be constructed from the combination of the on-shell type I supergravity action \eqref{AdS(1,1)aux} and the $\cN=2$ limit of the $\cN$-extended supersymmetric Lorentz Chern-Simons action \eqref{Stopo}, both evaluated in the unitary gauge \eqref{unitary}. The same is not true for the on-shell action \eqref{AdS(1,1)} due to the presence of non-algebraic contributions of the $\sU(1)$ gauge field in the topological sector. This was the main motivation for introducing the auxiliary vector field $A$ into the action \eqref{AdS(1,1)} by hand via the term \eqref{S_A}. 

However, a key inconsistency arises between the definitions of the gravitino field strengths used in the supergravity sector \eqref{AdS(1,1)aux} and those in the topological sector \eqref{Stopo}. Specifically there is no appearance of the $\sSO(2)\cong\sU(1)$ gauge field in the gravitino field strengths contained in the Rarita-Schwinger term of the supergravity action \eqref{AdS(1,1)aux}, whereas the gravitino field strengths used to construct the fermionic Chern-Simons term \eqref{SFCS} must contain the $\sU(1)$ connection by definition. The origin of this inconsistency comes from a difference in super-Poincar\'e algebras when one introduces the connection \eqref{superconn}. In particular, when constructing the $(1,1)$ AdS supergravity action, the $R$-symmetry subalgebra is absent for the choice of $p=q=1$, as this setting leads to $r_{IJ}=0$ in \eqref{SO(N) conn}. In contrast, constructing the $\cN=2$ supersymmetric Lorentz Chern-Simons action requires setting $p=2$ and $q=0$, which preserves a nontrivial $\sSO(2)$ connection. 

\subsubsection{Topologically massive type II supergravity}
Topologically massive type II supergravity arises from the coupling of the $\cN=2$ supersymmetric Lorentz Chern-Simons action \cite{RvN} with the off-shell $(2,0)$ AdS supergravity action \cite{KLRSTM,HIPT}. At this point we have only constructed the on-shell $(2,0)$ AdS supergravity action \eqref{AdS(2,0)}. If we want to couple to the $\cN=2$ supersymmetric Lorentz Chern-Simons action, we need the off-shell action for $(2,0)$ AdS supergravity since the topological sector contains non-algebraic contributions of the $\sU(1)$ gauge field. The off-shell action for $(2,0)$ AdS supergravity involves additional fields not supplied by our nonlinear realisation approach. In particular, there involves another $\sU(1)$ gauge field that naturally arises from the component reduction of the corresponding action superfunctional for off-shell type II supergravity \cite{KLRSTM}. While one may introduce this field into the action by hand, it would be somewhat artificial to our nonlinear realisation approach.\footnote{Introducing the $\sU(1)$ gauge field in the $(1,1)$ case was natural since our nonlinear realisation algebraic setup already contained a $\sU(1)$ gauge field in the connection \eqref{superconn}.} Furthermore, a calculation was performed after introducing this extra $\sU(1)$ gauge field and the variation of the action constructed did not vanish. 

\section{Conclusion}
In this paper, we have shown, using the nonlinear realisation approach to spontaneously broken supergravity in three dimensions developed in \cite{KS}, that the action \eqref{AdS(p,q)} constitutes a St\"uckelberg-type extension of the $(p,q)$ AdS supergravity theories proposed by Ach\'ucarro and Townsend. Under the same approach, we have described how to obtain the action \eqref{PoincareSUGRA}, which is also a St\"uckelberg-type extension of the $\cN$-extended Poincar\'e supergravity theory introduced by Marcus and Schwarz. Finally, by tweaking the algebraic setup of our approach, we derived a St\"uckelberg reformulation of the supersymmetric Lorentz Chern-Simons action for arbitrary $\cN$, given by the action \eqref{Stopo}. In the unitary gauge \eqref{unitary}, these three actions coincide with the standard ones given in the literature \cite{AT,MS,LR,NG}. They are gauge-fixed versions of the actions \eqref{AdS(p,q)}, \eqref{PoincareSUGRA} and \eqref{Stopo}.

Each of these actions is invariant under two different local $\cN$-extended supersymmetries. One of them acts on the Goldstini, while the other supersymmetry leaves the Goldstini inert. The $\cN$-extended Lorentz Chern-Simons action shares the former supersymmetry with the pure supergravity actions (Poincar\'e and AdS), but differs in the one that leaves the Goldstini inert. The supersymmetry that acts on the Goldstini can be used to gauge them away, and then the resulting actions turn into those given in the literature \cite{AT,MS,LR,NG}.

As pointed out in section \ref{Section4}, the action \eqref{Stopo} is functionally identical to the one for on-shell $\cN$-extended conformal supergravity \cite{LR,NG}, albeit not possessing invariance under dilatations and special conformal (bosonic and fermionic) transformations. It would be interesting to include these symmetries in a nonlinear realisation approach to on-shell $\cN$-extended conformal supergravity, which would presumably require beginning with the superconformal group. 

Extending beyond $\cN=1$ using the formalism of nonlinear realisations proved to reveal some limitations in our approach. Although the structure of $\cN$-extended Poincar\'e supergravity is uniquely determined by applying this formalism, one must deform the second local $\cN$-extended supersymmetry to include cosmological terms. This is no different to the $\cN=1$ case of AdS supergravity studied in \cite{KS}, where a deformation of the second local supersymmetry was performed. However, the key limitation arises when attempting to describe topologically massive supergravity theories for $\cN>1$. In particular, the $\cN=2$ extensions of topologically massive supergravity were incompatible with our nonlinear realisation approach. It seems that off-shell supergravity methods used in e.g. \cite{KLRSTM} appear natural for these considerations.

We conclude with a few final comments. As is well known, $\cN=1$ supergravity in 4D was discovered in 1976 by Ferrara, Freedman and van Nieuwenhuizen \cite{FvNF} and by Deser and Zumino \cite{DZ}.\footnote{A year later, a linearised supergravity action with auxiliary fields was constructed in \cite{AVS} by starting from a superfield approach. This was later rediscovered as alternative (new) minimal supergravity.} As discussed in section \ref{Section1}, it was demonstrated in \cite{K2021}, under the Volkov-Soroka framework, that the pure $\cN=1$ supergravity theory proposed in \cite{DZ} may be considered to be a special case of the $\cN=1$ Volkov-Soroka theory. An extension of this analysis of \cite{K2021} incorporating a cosmological term is given in appendix \ref{Appendix A}.\footnote{It was shown in \cite{KS} that the Volkov-Soroka approach restricted to the case of $\cN=0$ allows one to derive the action for gravity with a cosmological term.} In order to apply the Volkov-Soroka construction to $\cN=2$ supergravity, it seems necessary to introduce a central charge in the superalgebra and its associated vector gauge field. This will be studied elsewhere.
\\

\noindent
{\bf Acknowledgements:}\\
I am grateful to Sergei Kuzenko for suggesting the problem and for helpful discussions constructive to the manuscript. I also thank Ian McArthur for pointing out an important reference and for suggestions on the manuscript. Finally, I thank Igor Bandos and Dmitri Sorokin for their insightful comments. This work of JCS is supported by the Australian Government Research Training Program Scholarship.
\appendix
\section{4D $\cN=1$ supergravity with a cosmological term}\label{Appendix A}
A few years ago, it was shown \cite{K2021}, using a nonlinear realisation approach, that the action
\begin{align}
		S_{\text{SG}}&=S_{\text{EH}}+4S_{\text{RS}}\non\\
		&=\frac{1}{4}\int\ce_{abcd}E^a\wedge E^b\wedge R^{cd}+2\int\left(\Y\wedge E^a\wedge\s_a\cD\bar{\Y}-\cD\Y\wedge E^a\wedge\s_a\bar{\Y}\right)\label{4DSUGRA}
\end{align}
is invariant under two different local supersymmetries. The technical details of which are given in \cite{K2021}. The original Volkov-Soroka supersymmetry is described by the relations:
\begin{subequations}
	\begin{align}
		\d_\ce\Y^\a&=0~,\qquad \d_\ce E^a=0~,\\
		\d_\ce\W^{ab}&=0~,\\
		\d_\ce\Q^\a&=\e^\a~,\qquad \d_\ce X^a=\ii(\Q\s^a\bar{\e}-\e\s^a\bar{\Q})\label{A2c}~,\\
		\d_\ce\y^\a&=-\cD\e^\a~,\qquad \d_\ce e^a=2\ii(\y\s^a\bar{\e}-\e\s^a\bar{\y})\label{A2d}~.
	\end{align}
\end{subequations}
The second supersymmetry and the main original result of \cite{K2021} is defined by:
\begin{subequations}\label{4D2ndSUSY}
\begin{align}
	\d_{\bm \ce}\Y^\a&=-\cD\bm\e^\a~,\qquad \d_{\bm\ce} E^a=2\ii(\Y\s^a\bar{\bm\e}-\bm\e\s^a\bar{\Y})\label{A3a}~,\\
	\frac{1}{4}\ce_{abcd}\d_{\bm\ce}\W^{bc}\wedge E^d&=\bm\e\s_a\cD\bar{\Y}+\cD\Y\s_a\bar{\bm\e}\label{A3b}~,\\
		\d_{\bm\ce}\Q^\a&=0~,\qquad \d_{\bm\ce} X^a=0\label{A3c}~,\\
	\d_{\bm\ce}\y^\a&=-\cD\bm\e^\a+\Q^\b\d_{\bm\ce}\W_\b\,^\a~,\\
	\d_{\bm\ce}e^a&=2\ii(\Y\s^a\bar{\bm\e}-\bm\e\s^a\bar{\Y})+2\ii(\Q\s^a\cD\bar{\bm\e}-\cD\bm\e\s^a\bar{\Q})-\d_{\bm\ce}\W^a\,_bX^b\non\\
	&\quad+\frac{1}{2}\ce^{abcd}\d_{\bm\ce}\W_{bc}\Q\s_d\bar{\Q}~.
\end{align}
\end{subequations}
The latter supersymmetry is modelled on the former, and in particular the transformation laws \eqref{A3a} can be viewed as a natural generalisation of the Volkov-Soroka local supersymmetry \eqref{A2d}. 
The variation of the action \eqref{4DSUGRA} under the transformations \eqref{4D2ndSUSY} was calculated to be
\begin{equation}\label{4DSGvar}
	\d_{\bm\ce}S_{\rm SG}=-2\int\Big(\bm\e\s_a\cD\bar{\Y}+\cD\Y\s_a\bar{\bm\e}-\frac{1}{4}\ce_{abcd}\d_{\bm\ce}\W^{bc}\wedge E^d\Big)\wedge\dsT^a~.
\end{equation}
Indeed, this variation vanishes if $\d_{\bm\ce}\W$ is given by \eqref{A3b}.

Building on the previous results, we shall incorporate a supersymmetric cosmological term by first deforming the second supersymmetry \eqref{4D2ndSUSY} in the following way
	\begin{equation}
		\d_{\bm\ce}\Y^\a=-\cD\bm\e^\a-\l(\bar{\bm\e}\tilde{\s_a})^\a E^a\label{A5}~,
	\end{equation}
while keeping the $E^a$ and Goldstone field transformations the same, as given by the equations \eqref{A3a} and \eqref{A3c}. Here $\l$ is a constant real parameter. The elementary fields $\y^\a$ and $e^a$ each pick up an additional term proportional to $\l$:
\begin{subequations}\label{deformed}
	\begin{align}
		\d_{\bm\ce}\y^\a&=-\cD\bm\e^\a+\Q^\b\d_{\bm\ce}\W_\b\,^\a-\l(\bar{\bm\e}\tilde{\s}_a)^\a E^a\label{4Delem2ndSUSY1}~,\\
		\d_{\bm\ce}e^a&=2\ii(\Y\s^a\bar{\bm\e}-\bm\e\s^a\bar{\Y})+2\ii(\Q\s^a\cD\bar{\bm\e}-\cD\bm\e\s^a\bar{\Q})-\d_{\bm\ce}\W^a\,_bX^b\non\\
		&\quad+\frac{1}{2}\ce^{abcd}\d_{\bm\ce}\W_{bc}\Q\s_d\bar{\Q}+2\ii\l(\Q\s^a\tilde{\s}_b\bm\e-\bar{\bm\e}\tilde{\s}_b\s^a\bar{\Q})E^b\label{4Delem2ndSUSY2}~.
	\end{align}
\end{subequations}
As is now standard in these calculations, the dependence of $\d_{\bm\ce}\W$ in \eqref{4Delem2ndSUSY1} and \eqref{4Delem2ndSUSY2} is such that the composite fields $\Y^\a$ and $E^a$ remain unchanged upon varying the connection $\W\rightarrow\W+\d_{\bm\ce}\W$. As will be shown, the induced deformed second supersymmetry transformation law of $\W$ will be determined by demanding invariance under this new deformed local supersymmetry \eqref{A3a}, \eqref{A5}, \eqref{A3c} and \eqref{deformed}.

Next we add to the action \eqref{4DSUGRA} a supersymmetric cosmological term
\begin{align}\label{4Dsuper-cosm}
	S_{\text{super-cosm}}&=12\l^2S_{\text{cosm}}-16\ii\l S_{\text{mass}}\non\\
	&=\frac{\l^2}{2}\int\ce_{abcd}E^a\wedge E^b\wedge E^c\wedge E^d+4\l\int E^a\wedge E^b\left(\Y\wedge\s_{ab}\Y-\bar{\Y}\wedge\tilde{\s}_{ab}\bar{\Y}\right)~.
\end{align}

We now follow the approach used in \cite{K2021} by computing variations of the actions \eqref{4DSUGRA} and \eqref{4Dsuper-cosm} under the deformed second supersymmetry for the case $\bm\e\neq 0$ and $\bar{\bm\e}=0$ and adding the complex conjugate part at the end. The additional variation of the action \eqref{4DSUGRA} due to the term proportional to $\l$ in \eqref{A5} is given by
\begin{align}\label{addSGvar}
	\d_{\bm\e}^{(\l)}S_{\text{SG}}&=4\d_{\bm\e}^{(\l)}S_{\text{RS}}\non\\
	&=-2\l\int\Big(-4\Y\wedge\s_{ab}\cD\bm\e\wedge E^a\wedge E^b+2\Y\s_a\tilde{\s}_b\bm\e\wedge E^a\wedge\cD E^b\non\\
	&\quad-\Y\s_b\tilde{\s}_a\bm\e\wedge E^a\wedge\cD E^b\Big)~.
\end{align}
The total variation of the action \eqref{4Dsuper-cosm} under \eqref{A3a} and \eqref{A5} reads\footnote{There is no Lorentz connection variation contribution from this action.}
\begin{equation}\label{super-cosmvar}
	\d_{\bm\e}S_{\text{super-cosm}}=-16\ii\l\int E^a\wedge(\bm\e\Y)\wedge\Y\wedge\s_a\bar{\Y}-8\l\int\Y\wedge\s_{ab}\cD\bm\e\wedge E^a\wedge E^b~.
\end{equation}
Combining the variations \eqref{addSGvar} and \eqref{super-cosmvar} gives us
\begin{align}
		\d_{\bm\e}^{(\l)}S_{\text{SG}}+\d_{\bm\e}S_{\text{super-cosm}}&=-16\ii\l\int E^a\wedge(\bm\e\Y)\wedge\Y\wedge\s_a\bar{\Y}\non\\
		&\quad-2\l\int\Big(2\Y\s_a\tilde{\s}_b\bm\e\wedge E^a\wedge\cD E^b-\Y\s_b\tilde{\s}_a\bm\e\wedge E^a\wedge\cD E^b\Big)~.
\end{align}
The latter two terms can be algebraically simplified such that the first term is cancelled out and the variation is reduced to
\begin{equation}
		\d_{\bm\e}^{(\l)}S_{\text{SG}}+\d_{\bm\e}S_{\text{super-cosm}}=-2\int\Big(2\l\Y\s_b\tilde{\s}_a\bm\e\wedge E^b-\l\Y\s_a\tilde{\s}_b\bm\e\wedge E^b\Big)\wedge\dsT^a~.
\end{equation}
Adding the complex conjugate part for which $\bm\e=0$ and $\bar{\bm\e}\neq 0$ gives
\begin{align}\label{4Dlvar}
	\d_{\bm\ce}^{(\l)}S_{\text{SG}}+\d_{\bm\ce}S_{\text{super-cosm}}&=-2\int\Big(2\l\Y\s_b\tilde{\s}_a\bm\e-\l\Y\s_a\tilde{\s}_b\bm\e\non\\
	&\quad+2\l\bar{\bm\e}\tilde{\s}_a\s_b\bar{\Y}-\l\bar{\bm\e}\tilde{\s}_b\s_a\bar{\Y}\Big)\wedge E^b\wedge\dsT^a~.
\end{align}
Finally by summing the variations \eqref{4DSGvar} and \eqref{4Dlvar} we arrive at
\begin{align}
	\d_{\bm\ce}S&=-2\int\Big\{\bm\e\s_a\cD\bar{\Y}+\cD\Y\s_a\bar{\bm\e}-\frac{1}{4}\ce_{abcd}\d_{\bm\ce}\W^{bc}\wedge E^d\non\\
	&\quad+\l\big(2\Y\s_b\tilde{\s}_a\bm\e-\Y\s_a\tilde{\s}_b\bm\e\non\\
	&\quad+2\bar{\bm\e}\tilde{\s}_a\s_b\bar{\Y}-\bar{\bm\e}\tilde{\s}_b\s_a\bar{\Y}\big)\wedge E^b\Big\}\wedge\dsT^a~,
\end{align}
where we have denoted
\begin{equation}\label{4DAdSSUGRA}
	S=S_{\text{SG}}+S_{\text{super-cosm}}~.
\end{equation}
This action is invariant under the deformed local supersymmetry transformations provided
\begin{align}
	\frac{1}{4}\ce_{abcd}\d_{\bm\ce}\W^{bc}\wedge E^d&=\bm\e\s_a\cD\bar{\Y}+\cD\Y\s_a\bar{\bm\e}+\l\big(2\Y\s_b\tilde{\s}_a\bm\e-\Y\s_a\tilde{\s}_b\bm\e\non\\
	&\quad+2\bar{\bm\e}\tilde{\s}_a\s_b\bar{\Y}-\bar{\bm\e}\tilde{\s}_b\s_a\bar{\Y}\big)\wedge E^b~.
\end{align}

By construction, the action \eqref{4DAdSSUGRA} is invariant under gauge super-Poincar\'e transformations (see \cite{K2021} for the technical details). In particular, one can apply a local Poincar\'e translation along with the Volkov-Soroka local supersymmetry \eqref{A2c} to switch off the Goldstone fields $Z^A(x)=(X^a(x),\Q^\a(x),\bar{\Q}^{\dot{\a}}(x))$ by imposing the conditions
\begin{equation}
	X^a=0,\qquad \Q^\a=0~.
\end{equation}
As a result, the action \eqref{4DAdSSUGRA} turns into the supergravity action with a supersymmetric cosmological term proposed by Townsend \cite{T} in 1977.

\section{3D notation and conventions}\label{Appendix B}

In this appendix we collect key formulae of the 3D two-component spinor formalism that is described in \cite{KPTv}. The starting point for setting up this 3D spinor formalism is the 4D relativistic Pauli matrices
\begin{equation}
	(\s_{\underline{a}})_{\a\bd}:=(\id_2,\vec{\s})~,\qquad (\tilde{\s}_{\underline{a}})^{\ad\b}:=(\id_2,-\vec{\s})~,\qquad \underline{a}=0,1,2,3~,
\end{equation}
where $\vec{\s}=(\s_1,\s_2,\s_3)$ are the Pauli matrices. We remove the matrices with space index $\underline{a}=2$ and obtain the 3D gamma-matrices
\begin{subequations}
	\begin{align}\label{gammadown}
		(\s_{\underline{a}})_{\a\bd}\quad &\longrightarrow \quad (\g_a)_{\a\b}=(\g_a)_{\b\a}=(\id_2,\s_1,\s_3)~,\\
		(\tilde{\s}_{\underline{a}})^{\ad\b}\quad &\longrightarrow\quad(\g_a)^{\a\b}=(\g_a)^{\b\a}=\ce^{\a\g}\ce^{\b\d}(\g_a)_{\g\d}\label{gammaup}~,\quad a=0,1,2 ~.
	\end{align}
\end{subequations}

The $(\g_a)_{\a\b}$ and $(\g_a)^{\a\b}$ are invariant tensors of the Lorentz group $\sSO_0(2,1)$. They can be used to convert any three-vector 
$V^a$ into symmetric second-rank spinors
\begin{subequations}
	\bea
	\check{V} &=& (V_{\a\b})~, \qquad V_{\a\b} = V^a(\g_a)_{\a\b}~; \\
	\hat{V} &=& (V^{\a\b})~, \qquad V^{\a\b} = V^a(\g_a)^{\a\b}~.
	\label{B3.b}
	\eea
\end{subequations}
As is known, the invariance properties of $(\g_a)_{\a\b}$ and $(\g_a)^{\a\b}$ follow from 
the isomorphism $\sSO_0(2,1) \cong \sSL(2,\dsR) /{\mathbb Z}_2$ which is defined by associating with a group element $M \in \sSL(2, {\mathbb R})$ the linear transformation on the vector space of symmetric real $2\times 2$ matrices 
$\check{V} $
\bea
\check{V} \to M \check{V} M^{\rm T}~.
\eea

In the 3D case, the spinor indices are lowered  and raised using the $\sSL(2,\dsR)$ invariant spinor metric $\ce=(\ce_{\a\b}) = -(\ce_{\b \a}) $ and its inverse $\ce^{-1}=(\ce^{\a\b}) = - (\ce^{\b\a}) $, which are normalised by $\ce^{12}=-\ce_{12}=1$.
The rules for lowering and raising the spinor indices are:
\begin{equation}
	\y_\a=\ce_{\a\b}\y^\b~, \qquad
	\y^\a=\ce^{\a\b}\y_\b~.
\end{equation}
By construction, the $\g$-matrices (\ref{gammadown}) and (\ref{gammaup}) are real and symmetric. 

Properties of the 4D relativistic Pauli matrices imply analogous properties of the 3D $\gamma$-matrices. In particular, for the Dirac matrices
\begin{equation}\label{gammamat}
	\g_a:=\left( (\g_a)_\a\,^\b\right) =\ce^{\b\g}(\g_a)_{\a\g}=(-\ii\s_2,\s_3,-\s_1)
\end{equation}
we have the following identities
\begin{subequations}
	\begin{align}
		\g_a\g_b&=\h_{ab}\id_2+\ce_{abc}\g^c\label{prod}\quad \implies \quad 
		\{\g_a,\g_b\}=2\h_{ab}\id_2~,
		\\
		\g_a\g_b\g_c&=\h_{ab}\g_c-\h_{ac}\g_b+\h_{bc}\g_a+\ce_{abc}\id_2~,\\
		(\g^a)^{\a\b}(\g_a)^{\g\d}&=\ce^{\a\g}\ce^{\d\b}+\ce^{\a\d}\ce^{\g\b}~\label{gammacontraction},
	\end{align}
\end{subequations}
where the 3D Minkowski metric is $\h_{ab}=\h^{ab}=\text{diag}(-1,+1,+1)$, and the Levi-Civita tensors $\ce_{abc}$ and $\ce^{abc}$ are normalised by $\ce_{012}=-\ce^{012}=-1$. 

Throughout this paper, contractions of spinor indices are defined as follows:
\begin{subequations}
	\begin{align}
		\f\c&:=\f^\a\c_\a=\c\f~,\qquad \f^2:=\f\f~,\\
		\f\g_a\c&:=\f^\a(\g_a)_\a\,^\b\c_\b=-\c\g_a\f~.
	\end{align}
\end{subequations}
Here $\phi_\a$ and $\chi_\a$ are arbitrary anti-commuting spinors.

The Dirac matrices (\ref{gammamat}) along with the unit matrix, $\G_A:=\{\id_2,\g_a\}$, form a basis in the linear space of $2\times 2$ matrices. If we define the corresponding set with upper indices, $\G^A:=\{\id_2,\g^a\}$, we have the identity
\begin{equation}
	{\rm tr}\,(\G_A\G^B)=2\d_A\,^B~.
\end{equation}
In accordance with this identity, if $M=(M_\a\,^\b)$ and $N=(N_\a\,^\b)$ are $2\times 2$ matrices, then
\begin{subequations}
	\begin{align}\label{basis}
		M_\a\,^\b N_\g\,^\d&=\sum_{A}(C^A)_\a\,^\d(\G_A)_\g\,^\b~,\\
		(C^A)_\a\,^\d&=\frac{1}{2}M_\a\,^\b(\G^A)_\b\,^\g N_\g\,^\d~.\label{coeff}
	\end{align}
\end{subequations}
Now let $\y_1^\a$, $\y_2^\a$, $\y_3^\a$ and $\y_4^\a$ be arbitrary two-component spinors. Using the equations (\ref{basis}) and (\ref{coeff}) one can show that
\begin{equation}\label{Fierz}
	(\y_1M\y_2)(\y_3N\y_4)=-\frac{1}{2}(\y_1M\G^AN\y_4)(\y_3\G_A\y_2)~,
\end{equation}
which is the Fierz rearrangement rule for two-component spinors. 

The Levi-Civita tensor with lower curved-space indices, $\ce_{mnp}$, is defined by
\begin{equation}
	\ce_{mnp}=E\e_{mnp}=E_m\,^aE_n\,^bE_p\,^c\ce_{abc}~,
\end{equation}
where $E:=\det(E_m\,^a)$ and $\e_{mnp}$ is the Levi-Civita symbol. Its counterpart 
with upper curved-space indices $\ce^{mnp}$ is
\begin{equation}\label{upperLev}
	\ce^{mnp}=E^{-1}\e^{mnp}=E_a\,^mE_b\,^nE_c\,^p\ce^{abc}~.
\end{equation}

In three dimensions, any vector $F^a$ can be equivalently realised as a symmetric second-rank spinor 
$F_{\a\b}= F_{\b\a}$ or as an antisymmetric second-rank tensor $F_{ab} = -F_{ba}$. 
The former realisation is obtained using the gamma-matrices:
\begin{equation}
	F_{\a\b}:=(\g^a)_{\a\b}F_a=F_{\b\a}~,\qquad F^a=-\frac{1}{2}(\g^a)^{\a\b}F_{\a\b}~.
	\label{B.11}
\end{equation}
The antisymmetric tensor $F_{ab}$ is the Hodge-dual of $F_a$,
\begin{equation}
	F_{ab}=-\ce_{abc}F^c~, \qquad F_a=\frac{1}{2}\ce_{abc}F^{bc}~.
	\label{B.12}
\end{equation}
The symmetric spinor $F_{\a\b}$ is defined in terms of $F_{ab}$ as follows
\begin{equation}
	F_{\a\b}
	=\frac{1}{2}(\g^a)_{\a\b}\ce_{abc}F^{bc}~.
	\label{B.13}
\end{equation}
We emphasise that the three algebraic objects $F_a$, $F_{ab}$ and $F_{\a \b}$ 
are equivalent  to each other. The corresponding inner products are related to each other as follows:
\bea
-F^aG_a=
\hf F^{ab}G_{ab}=\hf F^{\a\b}G_{\a\b}
~.
\eea
More details can be found in \cite{KLT-M11}.

\begin{footnotesize}

\end{footnotesize}


\end{document}